\documentclass[article, twocolumn]{revtex4-1}

\usepackage{enumerate}
\usepackage{comment}
\usepackage{amsmath}
\usepackage{amssymb}
\usepackage{color}
\usepackage[dvipsnames]{xcolor}
\usepackage{tikz}
\usepackage{amsthm}
\usepackage{graphicx}
\usepackage{bbold}
\usepackage{empheq}

\newcommand{\be}{\begin{equation}}
\newcommand{\ee}{\end{equation}}
\newcommand{\bea}{\begin{eqnarray}}
\newcommand{\eea}{\end{eqnarray}}
\newcommand{\beb}{\begin{eqnarray*}}
\newcommand{\eeb}{\end{eqnarray*}}

\newcommand{\LD}{\langle}
\newcommand{\RD}{\rangle}

\newcommand{\eqn}{Eq.~}
\newcommand{\eqns}{Eqs.~}
\newcommand{\fig}{Fig.~}
\newcommand{\figs}{Figs.~}

\newcommand{\BLG}{BLG }
\newcommand{\LL}{LL  }
\newcommand{\LLs}{LLs }
\newcommand{\QPC}{QPC }

\begin{document}

\title{Influence of minivalleys and Berry curvature on electrostatically induced quantum wires in gapped bilayer graphene}

\author{Angelika Knothe}
\author{Vladimir Fal'ko}
\affiliation{National Graphene Institute, University of Manchester, Manchester M13 9PL, United Kingdom}

\date{\today}
\begin{abstract}
We show that the spectrum of subbands in an electrostatically defined quantum wire in gapped bilayer graphene (BLG) directly manifests the minivalley structure and reflects Berry curvature via the associated magnetic moment of the states in the low-energy bands of this two-dimensional material. We demonstrate how these appear in degeneracies of the low-energy minibands and their valley splitting, which develops linearly in a weak
magnetic field. Consequently, magneto-conductance of a ballistic point contact connecting two non-gapped areas of a bilayer through a gapped (top and bottom gated) barrier would reflect such degeneracies by the heights of the first few conductance steps developing upon the increase of the doping of the BLG conduction channel (we consider an adiabatic constriction, where conductance is set by the number of propagating ballistic modes in its narrowest part): $8e^2/h$ steps in a wide channel in BLG with a large gap, $4e^2/h$ steps in narrow channels, all splitting into a staircase of $2e^2/h$ steps upon lifting valley degeneracy by a magnetic field.        
\end{abstract}
\maketitle

\section{Introduction}
\begin{figure}[h!]
\includegraphics[width=0.4\textwidth]{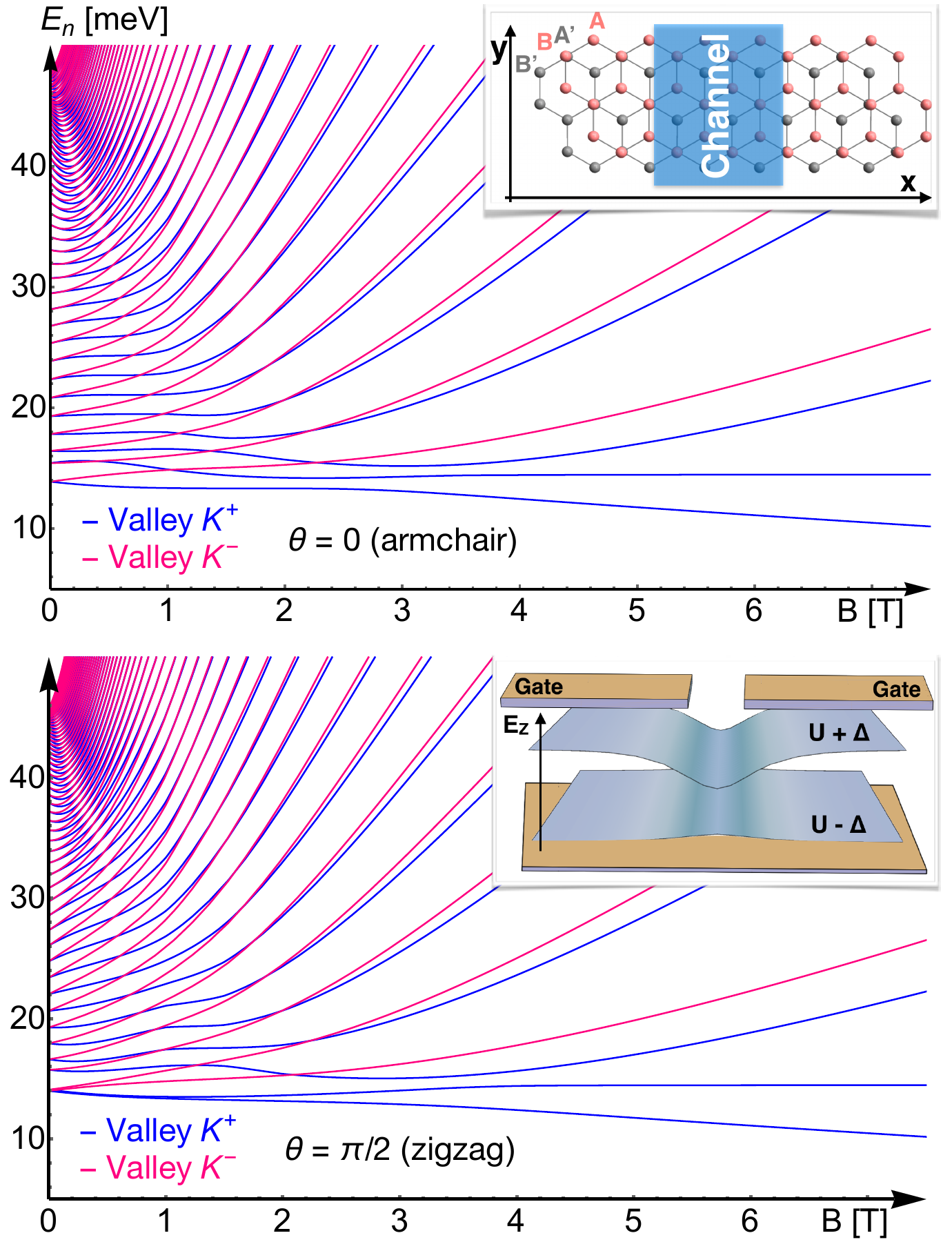}
 \caption{Subband edges $E_n$ (conduction band) in the electron channels in a \BLG quantum wire as a function of an external magnetic field, B, for $K^\pm$ valleys. The device parameters (explained later in the text) are $\Delta_0=100$ meV, $L=85$ nm, $U_0=-20$ meV, and $\beta=0.3$. Top/bottom panels correspond to armchair/zigzag channel direction.  Upper inset: sketch of the channel geometry (blue region), where A, B, A', B' label the basis atoms of the \BLG lattice. Lower inset: The device architecture where voltage applied to the bottom gate and top split gates determine the confinement potential $U(x)$ and  the modulated gap $\Delta(x)$ forming the channel.}
\label{fig:EwB_D100_zoom}
\end{figure}

The development of hardware for quantum technology applications requires materials where an operation of a qubit, such as a spin state of an electron in a quantum dot \cite{Loss1998},
is not hindered by spin decoherence due to its interaction with environment. In conventional semiconductors, hyperfine interaction of electron's spin with nuclear spins
appears to lead to an unbeatable decoherence \cite{Fischer2009, Yao2006, Klauser2007}, leading to a proposal \cite{Rycerz2007, Trauzettel2007} to employ electron's spin
and valley degrees of freedom in quantum circuits fabricated from graphene with spinless C12 nuclei. While quantum dot qubits and quantum wire readouts \cite{BischoffDominik2015}
fabricated from monolayer graphene suffer from disorder caused by functionalisation of their edges, the use of bilayer graphene (BLG) \cite{McCann2006}, where a substantial band gap
($\sim 200$ meV) can be induced by electrostatic gating \cite{Heersche2007, Varlet2014, Varlet2015} has been suggested \cite{Falko2007} as a possible route towards creating
quantum circuits with a long spin-coherence time. Recently, electrostatically controlled quantum wires \cite{Droscher2012, Overweg2018, Kraft2018, Hunt2017} and even dots \cite{Allen2012, Goossens2012, Eich2018} in BLG have been successfully fabricated and operated in the Coulomb blockade (for dots) and ballistic conduction (for wires) regimes.
In such devices,  similar to those sketched in \fig\ref{fig:EwB_D100_zoom}, dual (top and bottom) gating \cite{Yan2010, Avsar2016} permit to control both interlayer asymmetry gap, $\Delta$, generated by  a displacement field, $E_Z$,
and the Fermi level in the conduction channel.
\begin{figure}[h]
 \centering
\includegraphics[width=0.4\textwidth]{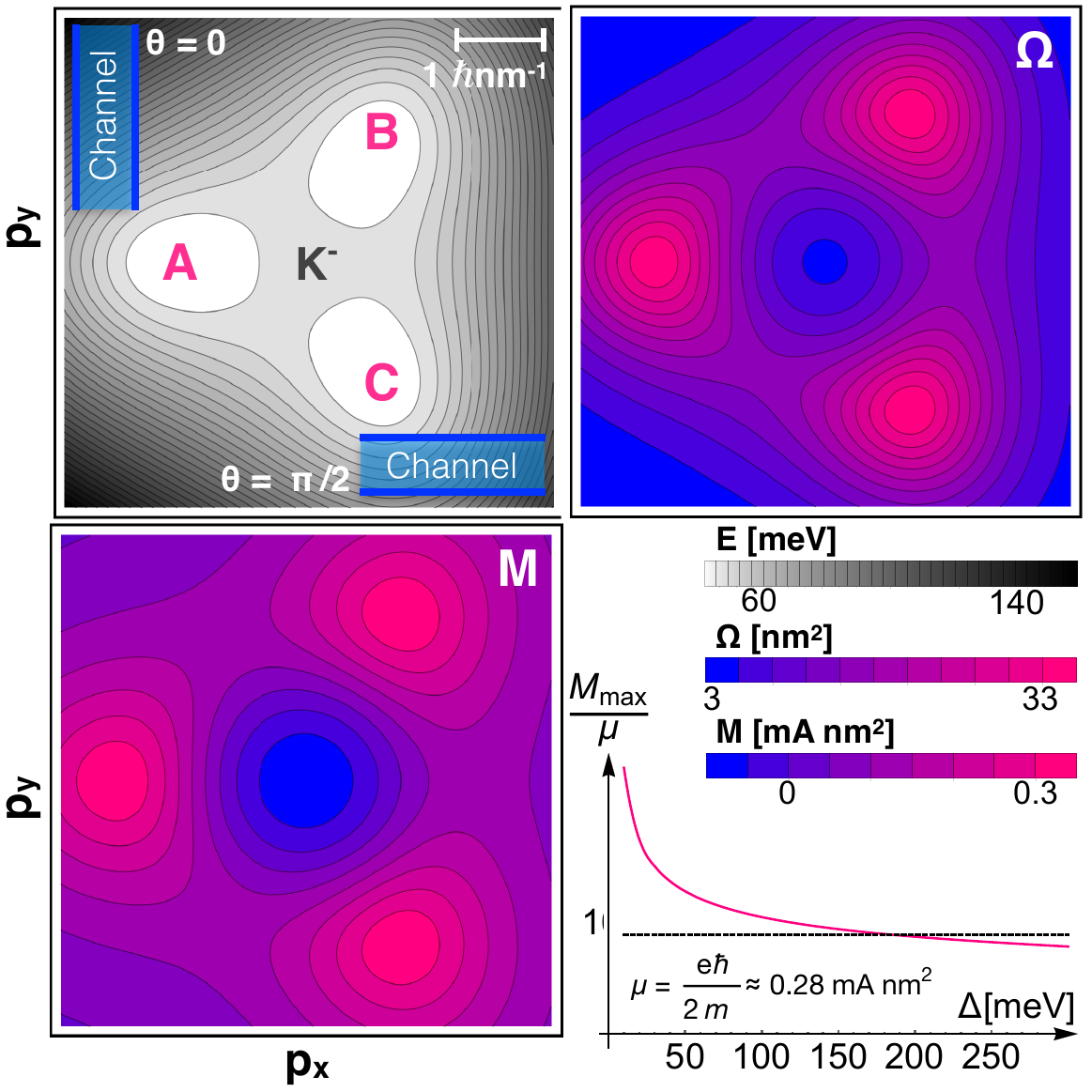}
 \caption{Contour energy plot for the 2D  dispersion of homogeneous gapped BLG's lowest conduction band for $\Delta=100$ meV (grayscale), with minivalleys A, B, and C in valley $K^-$. Blue bars indicate the orientation of the wire axis along the armchair  ($\theta=0$) or zigzag  ($\theta=\pi/2$) direction in graphene. The corresponding Berry curvature, $\Omega$, and orbital magnetic moment, $M$, are shown in the top right and bottom left panel, respectively. The bottom right panel shows the $\Delta$-dependence of the maximum value of such magnetic moment, $M_{max}$, carried by a plane wave state of electrons with momentum $\mathbf{p}$. }
\label{fig:Berry}
\end{figure}

 Here, we present a detailed theoretical analysis of electronic properties of ballistic quantum wires and their magnetotransport characteristics. We calculate the dispersions
of the one-dimensional (1D) modes, $E_n(k)$, in the wire, which appear to reflect the formation of a triplet of minivalleys around both K$^+$ and K$^-$ Brillouin zone corners
of the spectrum of BLG upon opening its interlayer asymmetry gap, $\Delta$, using the vertical displacement field. We find that, for wide channels and large gaps $\Delta$,
these minivalleys set an approximate degeneracy of the edges of the lowest 1D subbands in the electrostatically defined 1D channel in BLG, as well as a dependence of the subband
spectra on the crystallographic orientation of the channel axis. We also find that the Berry curvature of electron states in the gapped BLG minivalleys, and  the associated magnetisation of the electron
states \cite{Xiao2010, Chang1996a}, lead to the linear in magnetic field splitting of the valley degeneracy of the subbands. All these features are illustrated
in \fig\ref{fig:EwB_D100_zoom}, where we plot the energies of the dispersion minima, $E_n(k_{min})$, for both K$^+$ and K$^-$ valleys in a BLG channel. After taking into account spin degeneracy,
the subband edge spectra in \fig\ref{fig:EwB_D100_zoom} can be used to predict the staircase of conductance steps forming upon filling ballistic wires with carriers: crossing each of the levels plotted by the Fermi
energy corresponds to an $2e^2/h$ conductance step. From this, one can see that, at $B=0$, a wire with the axis aligned with the armchair direction of the graphene lattice would have all steps with
the height of $4e^2/h$, whereas a wire aligned along the zigzag direction would feature a twice-higher first step, $8e^2/h$.

The above described properties of quantum wires in gapped BLG have been determined using the four-band BLG Hamiltonian \cite{McCann2007, McCann2013}
\begin{widetext}
\begin{equation}
 H^{\xi}_{BLG}=\xi
\setlength{\arraycolsep}{-2pt} \begin{pmatrix} 
\xi U(x)-\frac{1}{2}\Delta(x) & v_3\pi & 0 &v \pi^{\dagger}\\
 v_3 \pi^{\dagger}&\xi U(x)+\frac{1}{2}\Delta(x) & v\pi &0\\
 0 & v\pi^{\dagger} & \xi U(x)+\frac{1}{2}\Delta(x) & \xi \gamma_1\\
 v\pi & 0 & \xi \gamma_1 &\xi U(x)-\frac{1}{2}\Delta(x)
\end{pmatrix},
\begin{Bmatrix} 
\pi=p_x+ip_y,\,  \pi^{\dagger}=p_x-ip_y,\\
\text{with }\mathbf{p}=-i\hbar\nabla-\frac{e}{c}\mathbf{A}, \\
v=1.02*10^6 \text{ m/s},\\
  v_3\approx0.12 v,\,\gamma_1=38\text{ eV},
\end{Bmatrix}
\label{eqn:H}
\end{equation}
\end{widetext}
 written in the basis $\Phi_{K^+}=(\Psi_{A},\Psi_{B^{\prime}},\Psi_{A^{\prime}},\Psi_{B})$ or $\Phi_{K^-}=(\Psi_{B^{\prime}},\Psi_{A},\Psi_{B},\Psi_{A^{\prime}})$ of states on the four \BLG sub-lattices sketched in the two valleys, $K^{\pm}$ (for $\xi=\pm1$). The diagonal terms  account for the spatially modulated confinement potential, $U(x)$ and an electrostatically modulated gap, $\Delta(x)$, chosen in the  form 
\begin{equation}
U(x)=\frac{U_0}{\cosh{\frac{x}{L}}} \; , \, \Delta(x)=\Delta_0\big[1-\frac{\beta }{\cosh{\frac{x}{L}}}\big].
\label{eqn:UD}
\end{equation}
 The choice of $U(x)$ and $\Delta(x)$ is motivated by recent experiments \cite{Overweg2018}, where  simulations have been performed to estimate the electrostatic potential profile inside the channel. For homogenous \BLG with an interlayer asymmetry gap $\Delta$   features  \cite{McCann2007} four valley degenerate bands, $\pm E_{\alpha=1,2}$, such that
\begin{align}
\nonumber E_{\alpha}^2&=    \frac{\gamma_1^2}{2} +\frac{\Delta^2}{4} +(1+\frac{\tau^2}{2})\frac{\gamma_1 p^2}{2m} \\
\nonumber&+ (-1)^{\alpha}v^2\Bigg( \frac{(2\gamma_1 m-\tau^2p^2 )^2}{4}  +4\xi mv^3\tau p^3 \cos3\varphi   \\
& +p^2 [2m\gamma_1 +\frac{\Delta^2}{v^2} +\tau^2p^2]    \Bigg)^{\frac{1}{2}}.
\end{align}
Here, $\mathbf{p}=p(\cos\varphi,\sin\varphi)$ is the momentum near the $K$ point, $m\approx\frac{\gamma_1}{2v^2}\approx0.032m_e$ is the effective mass of electrons in gapless BLG, and $\tau=\frac{v_3}{v}\approx0.1$ parametrizes skew interlayer hopping \cite{McCann2006, Varlet2015}. The dispersion of the lower conduction band ($\alpha=1$) in the $K^{-}$ valley is plotted in \fig\ref{fig:Berry} (for $\Delta = 100$ meV).

The dispersion features three minivalleys around each $K$ point \cite{Varlet2014, Varlet2015}. The corresponding bands carry Berry curvature, $ \Omega(\mathbf{p})$, and orbital magnetic moment, $\mathbf{M}(\mathbf{p})=M(\mathbf{p})\mathbf{e}_z$, determined by the Bloch functions as \cite{Xiao2010, Chang1996a},
\begin{align}
\nonumber{\Omega}&=i\hbar^2\LD\mathbf{\nabla}_{\mathbf{p}}\Phi(\mathbf{p})|\times|\mathbf{\nabla}_{\mathbf{p}}\Phi(\mathbf{p})\RD\cdot\mathbf{e}_z ,\\
{M}&=-i{e}{\hbar}\LD\mathbf{\nabla}_{\mathbf{p}}\Phi(\mathbf{p})|\times [\epsilon(\mathbf{p})-H(\mathbf{p})] |\mathbf{\nabla}_{\mathbf{p}}\Phi(\mathbf{p})\RD\cdot\mathbf{e}_z.
\label{eqn:Berry}
\end{align}
Here, $\mathbf{\nabla}_{\mathbf{p}}=(\partial_{p_x},\partial_{p_y})$, $"\times"$ is the cross product,  $\epsilon(\mathbf{p})$ is the band energy, and $e>0$. Both $\Omega$ and $M$, computed numerically from the four-band model of BLG, are shown in  \fig\ref{fig:Berry} for the lower conduction band in the $K^-$ valley (the sign of $\Omega$ and $M$ is reversed in the $K^+$ valley). They exhibit maxima at the minivalleys, while $M(0)\approx\Omega(0)\approx 0$. Note that the two-band model of \BLG  \cite{McCann2006, McCann2007} with $\tau\rightarrow0$ \cite{Park2017, Fuchs2010}, for which $\epsilon(\mathbf{p})=\sqrt{(\frac{p^2}{2m})^2+(\frac{\Delta}{2})^2}$, gives  decent estimates for $\Omega\approx-\xi\frac{\hbar^2}{2m}\frac{p^2}{2m}\frac{\Delta}{\epsilon(\mathbf{p})^3}$ and $M\approx  \epsilon(\mathbf{p}) \frac{e}{\hbar}\Omega=-\xi\frac{e\hbar}{2m}\frac{p^2}{2m}\frac{\Delta}{\epsilon(\mathbf{p})^2}$   (see Sec.~\ref{sec:appBerry}). 

\section{Manifestation of gapped BLG minivalleys and Berry curvature in the spectral properties of electrostatically controlled quantum wires}

The electronic properties of gapped \BLG also determine the features of the dispersion of an electrostatically induced channel. The states close to the edges of the minibands forming in the 1D \BLG channel are determined by the electron states of the minivalleys of gapped \BLG . The quantisation of the electron motion perpendicular to the channel axis reflects the anisotropy of the effective mass in the minivalleys, which, in the limit of small, but finite gaps ($\delta=\frac{\Delta}{\gamma_1}\ll1$ but $\Delta>v_3^2 m$) and small interlayer hopping ($\tau\ll1$) can be found  approximately to read (see Sec.~\ref{sec:appMass}),
\begin{align}
\nonumber&m/m_{p}^{-1}\approx\frac{1}{8\delta} [8\delta^4+8\delta^2+\tau(3\sqrt{8\delta^2+\tau^2}+\tau)]  ,\\
&m/m_{\varphi}^{-1}\approx\frac{9}{8\delta}  \tau(\sqrt{8\delta^2+\tau^2}+3\tau)  ,
\end{align}
where $\Delta=(1-\beta)\Delta_0$ is the gap in the middle of the channel. This determines the dependence of the miniband spectrum on the orientation of the channel axis (blue bars in \fig\ref{fig:Berry}). Moreover, the non-trivial topological properties of the \BLG bands (Berry curvature and corresponding  magnetic moment) determine the response to an external magnetic field, $B$. For a weak magnetic field, this leads to a valley dependent shift, 
\begin{equation}
\delta\epsilon_{\xi}=-MB=-\xi |M|B, 
\label{eqn:lin}
\end{equation}
of the bottom miniband energy, which leads to the valley splitting of the miniband spectra. At high magnetic fields this valley splitting culminates in a full valley polarization of the lowest two minibands resulting from the sublattice polarisation of the "zero-energy" $(n=0,1)$ Landau levels (LLs) in \BLG \cite{Trauzettel2007}.

To develop a detailed qualitative description of the spectra of the 1D channel in BLG, we diagonalize the Hamiltonian in \eqn\eqref{eqn:H} numerically \cite{Varlet2014, Varlet2015} in a basis,
\begin{equation} 
\begin{pmatrix}
\psi_n(\tilde{x})\\
0\\
0\\
0
\end{pmatrix},\;
\begin{pmatrix}
0\\
\psi_n(\tilde{x})\\
0\\
0
\end{pmatrix},\;
\begin{pmatrix}
0\\
0\\
\psi_n(\tilde{x})\\
0
\end{pmatrix},\;
\begin{pmatrix}
0\\
0\\
0\\
\psi_n(\tilde{x})
\end{pmatrix},
\label{eqn:Basis}
\end{equation}

where $\psi_n(\tilde{x})=\sqrt{\frac{\alpha}{\sqrt{\pi}2^n n!}}\; e^{-\frac{1}{2}(\alpha \tilde{x})^2}\mathcal{H}_n(\alpha \tilde{x})$ are harmonic oscillator functions with a scaling factor $\alpha$    adapted to the width, $L$, of the quantum well. We assume free propagation of the electrons along the channel axis $\tilde{y}=x\sin\theta+y\cos\theta$, and quantization across the channel axis $\tilde{x}=x\cos\theta-y\sin\theta$: $\Psi_n(\tilde{\mathbf{r}})=e^{ik_y \tilde{y} }\Phi_{n}(\tilde{x})$.

 The wave vector $k$ is related to the momentum $p$ by $\hbar$. For every set of system parameters we use \eqn\eqref{eqn:H} and this basis to compute the Hamiltonian matrix and diagonalize it to obtain the energy spectrum for each $k_y$ point. Convergence is reached when the energy levels do not change anymore upon including a higher number of basis states.\\

\begin{figure}[h!]
 \centering
\includegraphics[width=0.49\textwidth]{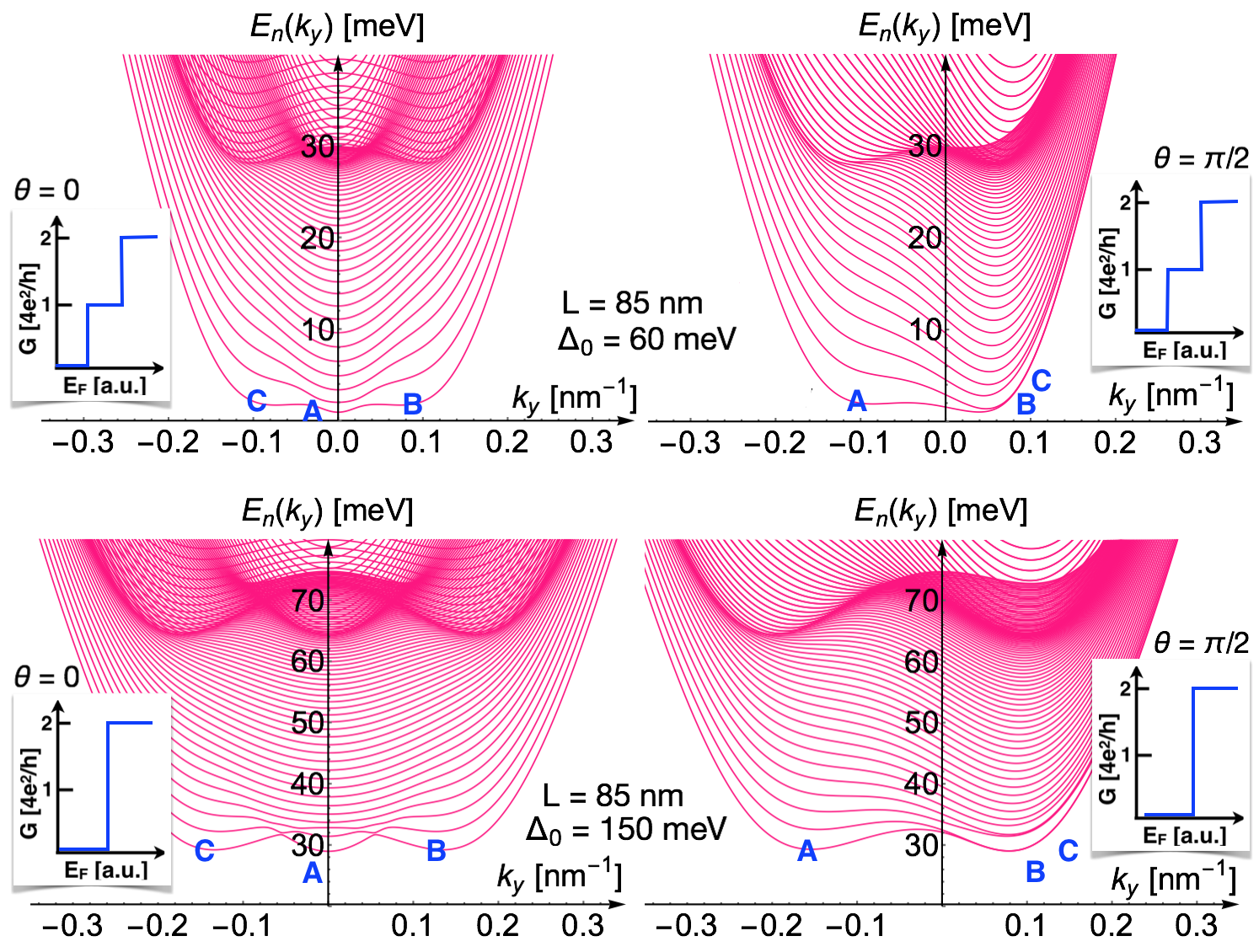}
 \caption{Energy levels of the conduction band for \BLG in the presence of a confinement potential $U(x)$ and  a modulated gap $\Delta(x)$ as given by \eqn\eqref{eqn:UD} for several values of $\Delta_0$ and  angle $\theta$. The marks A, B, C refer to the minivalleys indicated in \fig\ref{fig:Berry}.  The insets illustrate how the spectra convert into conductance steps $G(E_F)$ upon filling the quantum wire with electrons.  
 }
\label{DispB0L50D60a250}
\end{figure}

To quantify the channel spectra, we perform a numerical analysis of the Hamiltonian given in \eqn\eqref{eqn:H} for several values of $\Delta_0$ and $L$, with and without magnetic field, and for 1D channels oriented along the armchair ($\theta=0$) or the zigzag ($\theta=\pi/2$) direction. In \fig\ref{DispB0L50D60a250} we show the spectra of minibands in the channel with  $L=85$ nm and  for different values of $\Delta_0$ at $B=0$. Further examples of spectra demonstrating the influence of the various model parameters are provided in Sec.\ref{sec:app}.

Adiabatically, only the outer modes at non-zero momentum, indexed by $B$ and $C$ in \fig\ref{DispB0L50D60a250}, would be populated by non-equilibrium electrons driven by a bias voltage applied to the bulk electrodes at the end of the 1D channel structure. Therefore, the fact that some of the calculated dispersions feature a band inversion  does not affect the mode counting (for details, see Sec.~\ref{sec:appPop}). From counting the number of current-carrying modes provided by the lowest-energy band, we conjecture the conductance step at the band-edge shown in sketches in \fig\ref{DispB0L50D60a250}. For narrow channels and/or smaller $\Delta_0$, the lowest energy levels are well separated, resulting in the first conductance step of 4 ${e^2}/{h}$ that should happen upon filling the channel with electrons (the factor of 4 in the height of the step reflects the spin and valley degeneracy). For wider channels and/or larger $\Delta_0$, the lowest-energy bands merge together (in fact, this is an almost-degeneracy with an exponentially small splitting of the band edges). This results in the step of 8 ${e^2}/{h} $ at the conductance threshold. Note that this simple mode counting picture does not account neither for the influence of disorder in the system nor for potential relaxation of the non-equilibrium electrons via scattering (see Sec.~\ref{sec:appPop}). Their effect will be subject of further studies.

\begin{figure}[t!]
 \centering
\includegraphics[width=0.5\textwidth]{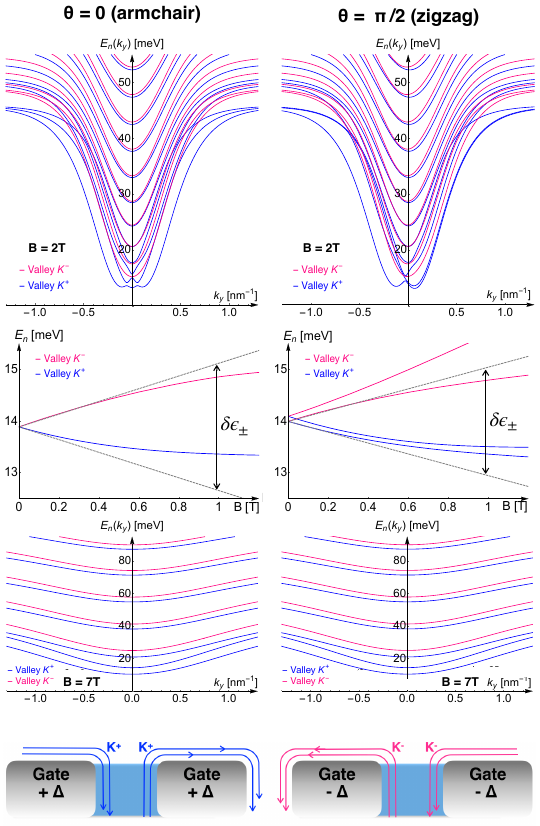}
 \caption{Subband spectra in a BLG wire in a magnetic field ($\Delta_0=100$ meV, $L=85 $ nm). Small-field regime (top/second row) displays a linear in B valley splitting (gray asymptotes in the second row panels) expected from \eqns\ref{eqn:Berry} and \ref{eqn:lin}. High-field regime (third/bottom row): electronic channels  are in fact edge states of the lowest LLs in gapped BLG. In   structures with multiple split gates and variable sign of the gap (bottom panel) the states in the opposite valleys are realized in the parts with opposite sign of the gaps and come together at the $\pm\Delta$ interface where they form topologically protected channels \cite{Martin2008, Cosma2015}.}
\label{fig:approx}
\end{figure}
To take into account a magnetic field, in \eqn\eqref{eqn:H} we use the vector potential in Landau gauge along the channel: $\mathbf{A}=B(y\sin\theta,-x\cos\theta,0).$ In the presence of a gap and a non-zero magnetic field both spatial and time reversal symmetry are broken. Therefore, for $B\neq0$, we compute the spectra for the two valleys $K^\pm$, separately.  Examples of such spectra for $\Delta_0=100$ meV, $L=85$ nm, and $B=2$ T are shown in \fig\ref{fig:approx}. The lowest bands in these spectra clearly displays $K^{\pm}$ valley splitting, for which the difference $\delta\epsilon_{\pm}=\delta\epsilon_{+}-\delta\epsilon_{-}$ of the lower subband edges is described very well by the Zeeman splitting of magnetic-moment carrying states in the minivalleys of gapped \BLG, 
\begin{equation}
\delta\epsilon_{\pm} \approx -2 | M |B.
\label{eqn:linear}
\end{equation}
with the same slope but opposite signs in the valleys $K^+$ and $K^-$. We demonstrate the agreement of this linear behaviour in $B$ with the numerics at small magnetic fields in \fig\ref{fig:approx}.
 In the  limit of large magnetic fields, the subbands in the channel become spatially modulated and evolve into the \LLs of BLG. For \LL number $n$ these are given approximately by \cite{McCann2006, McCann2013}
\begin{align}
\nonumber&E_0\approx  E_{B=0,  0} \;\text{     and      } \; E_1\approx  E_{B=0,  0}  - \xi\frac{\Delta}{\gamma_1}\hbar\omega_c,\\
&E_{n,\xi}\approx E_{B=0, n-1}+\hbar\omega_c\sqrt{n(n-1)}-\xi\frac{\Delta}{2\gamma_1}\hbar\omega_c,
\label{eqn:scaleB1}
\end{align}
where $\omega_c=eB/m$. With $E_{B=0,n}$ we denote the $n$th $B=0$ conduction band level  and  $\Delta$ is the  gap at the centre of the channel [for the gap profile in \eqn\eqref{eqn:H}, $\Delta=(1-\beta)\Delta_0$]. For the $n=0,1$ \LLs 
  the electron valley index is linked to the sublattice so that the two lowest minibands on the conduction band side of the BLG spectrum are always in one valley, as represented by the colours in \fig\ref{fig:EwB_D100_zoom}. Note that the states in the opposite valley are buried in the valence band, hence, excluded from the transport of the channel. To create a channel with  electrons based on the opposite BLG valley ($K^-$ instead of $K^+$) one would need to either invert the magnetic field, or invert the sign of the gap (by inverting the gate voltage), as sketched in the bottom image of \fig\ref{fig:approx}.


\section{Evolution of BLG quantum wire SPECTRA upon variation of the BLG asymmetry gap, crystallographic orientation of the channel, and magnetic field}



\subsection{Examples of spectra over a large parameter range}
\label{sec:app}

\begin{figure}[h!]
 \centering
\includegraphics[width=0.49\textwidth]{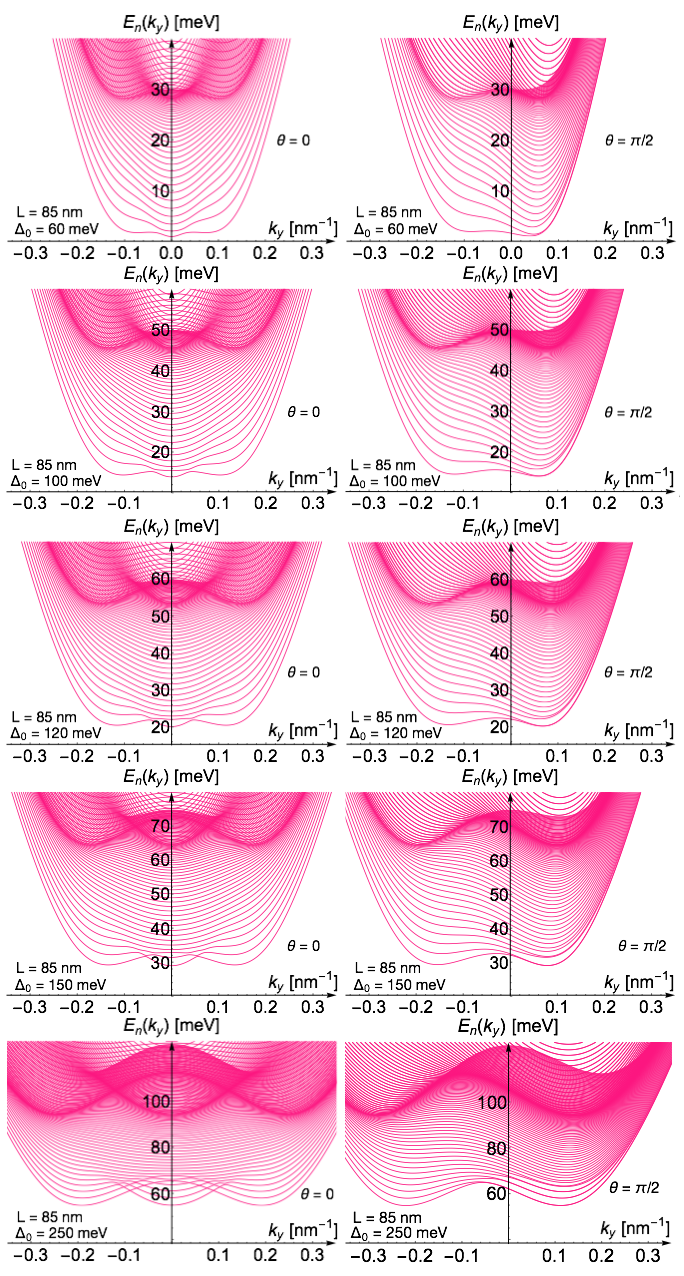}
 \caption{Energy levels of the conduction band for the \BLG \QPC without magnetic field for different values of  $\Delta_0$ for orientation $\theta=0$ in the left column and $\theta=\pi/2$ in the right column. }
\label{fig:QPC_SpectraNoB}
\end{figure}

\begin{figure}[h]
 \centering
\includegraphics[width=0.49\textwidth]{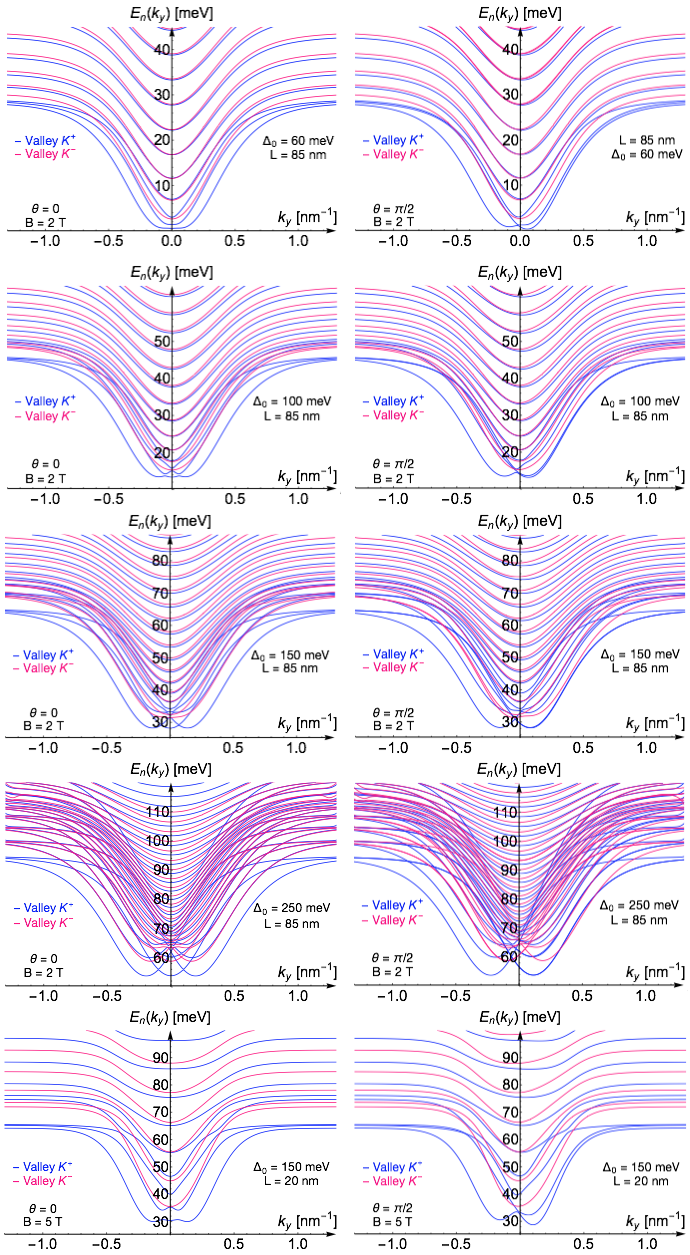}
 \caption{Channel spectra (conduction band) for the \BLG \QPC in the presence of a magnetic field for different values  of $\Delta_0$ and  $L$. }
\label{fig:QPC_SpectraWB}
\end{figure}

\begin{figure}[h]
 \centering
\includegraphics[width=0.49\textwidth]{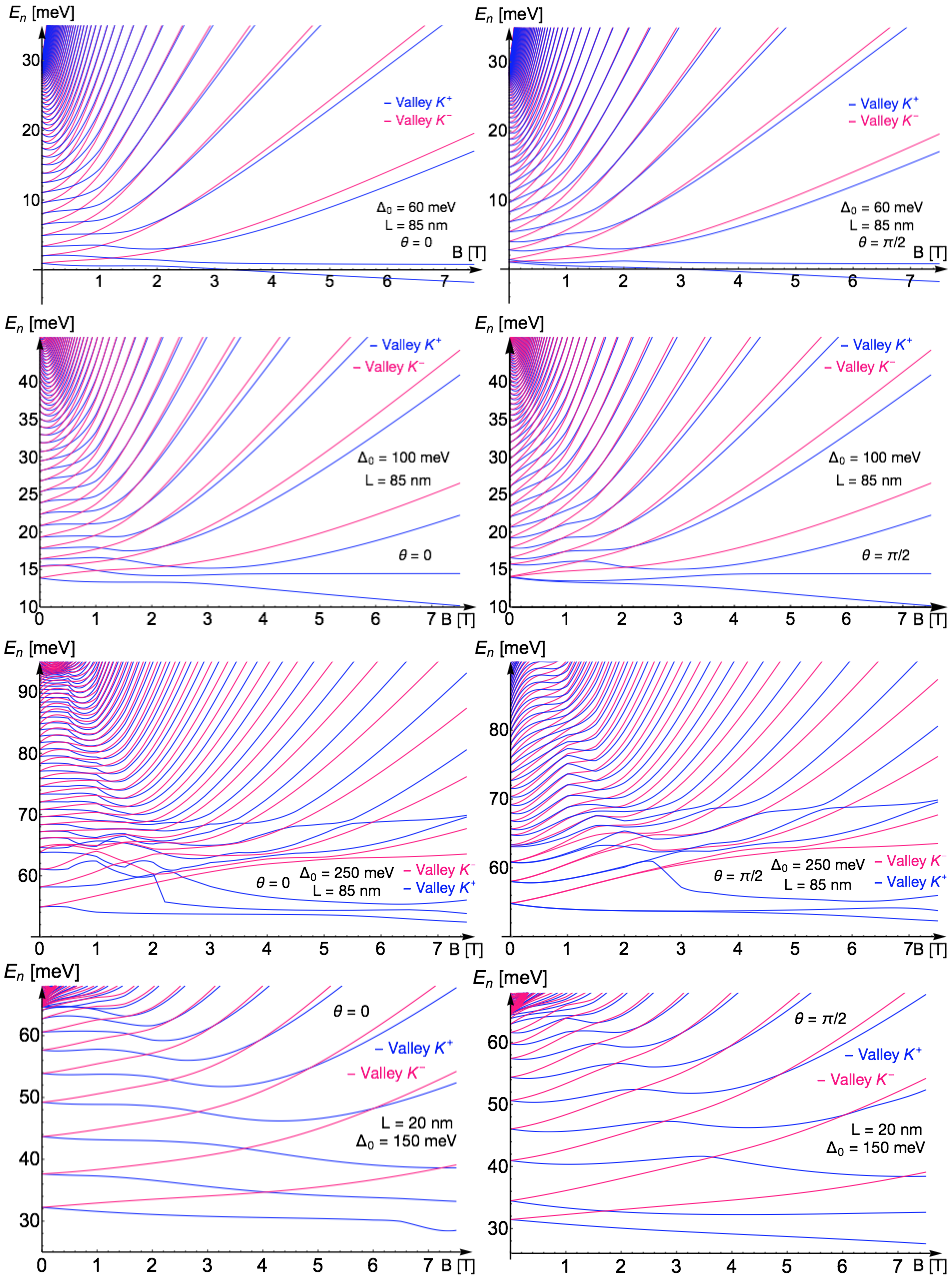}
 \caption{Energy levels of the conduction band for the \BLG \QPC as a function of $B$ for different values of  $\Delta_0$ and $L$.}
\label{fig:QPC_EwBTrigo}
\end{figure}

We discuss the dependence of the channel spectra on the various system parameters. Figure \ref{fig:QPC_SpectraNoB} demonstrates the evolution of the non-magnetic conduction band levels with increasing $\Delta_0$  for $L = 85$ nm for orientation angle $\theta=0$ in the left column and $\theta=\pi/2$ in the right column. There is a set of sizequantized energy levels as a function of the transverse momentum $k_y$, before, above a certain energy, the continuous spectrum is reached. The lowest conduction band edge develops a structure with multiple minima with increasing $\Delta_0$. At a certain critical value of $\Delta_0$ (which depends on $L$ and on the orientation), the spectra exhibit an additional degeneracy when the lowest two energy levels touch. Figure \ref{fig:QPC_SpectraWB} shows corresponding channel spectra when a non-zero external magnetic field is considered. In the presence of a magnetic field the spectra form LLs. Breaking of both spatial inversion symmetry and time reversal symmetry entails valley symmetry breaking and therefore we obtain two unrelated spectra  for the valleys $K^+$ and $K^-$. For the case without confinement potential, \textit{i.e.}, $U\equiv0$, particle hole symmetry in one valley is broken in a non-zero magnetic field, but is restored when both valleys are included. Figure \ref{fig:QPC_SpectraWB} compares the \LLs in the conduction band for different values of $\Delta_0$,  $L $, and the field strength $B$ for both channel orientations $\theta=0$ (left column) and  $\theta=\pi/2$ (right column).  We observe a similar multi-minima structure as in the zero-field case (where, additionally, symmetry between the two valleys is broken) before, above a certain threshold, nearly flat \LLs are formed. The  splittings of the subband edges in the 1D channel which result from the valley splitting lead to the splittings of the conduction quantization steps. In \fig\ref{fig:QPC_EwBTrigo} we show additional examples for the lower conduction band edges as a function of the magnetic field strength for different  $\Delta_0$ and different $L$. We analyse the discrete energy spectra as a function of $B$ from low values of $B$, where the behaviour of the levels and the gaps is dictated by the confinement, to large magnetic fields where the spectra evolve into the $B$-field driven \LL spectra. At zero magnetic field the level splitting of the lowest energy levels scales as $1/L$, and is almost independent of $\Delta_0$. Conversely, the scaling of the \LLs at large magnetic fields is governed by   $\Delta_0$, see \eqn\ref{eqn:scaleB1}. Also, with increasing $\Delta_0$, additional features in the spectra appear, notably for the lowest conduction band levels. Therefore, we find the magnetic field positions of the level crossings in  \fig\ref{fig:QPC_EwBTrigo} to  depend on both parameters, $L$ and $\Delta_0$. For all choices of the system parameters, above a certain value of the magnetic field, the two lowest energy levels stem from the same valley.  As the two zero-energy states of the $K^-$-valley are buried in the valence band, we observe this pairing between states $N+1$ and $N-1$ in the large field limit, where $N$ labels the energy levels at zero magnetic field. As a consequence, the two lowest energy states in this regime always stem from the same valley (see \fig\ref{fig:EwB_D100_zoom}). This allows for the postulation of  valley polarized currents in a \BLG quantum point contact (QPC) in the presence of a magnetic field. The possibility of creating valley polarized currents in this structure is hence a robust feature of this structure in this 2D material and does not require a particular choice of system parameters. 

\subsection{Berry curvature and Magnetic Moment}
\label{sec:appBerry}

\begin{figure*}[t!]
 \centering
\includegraphics[width=0.9\textwidth]{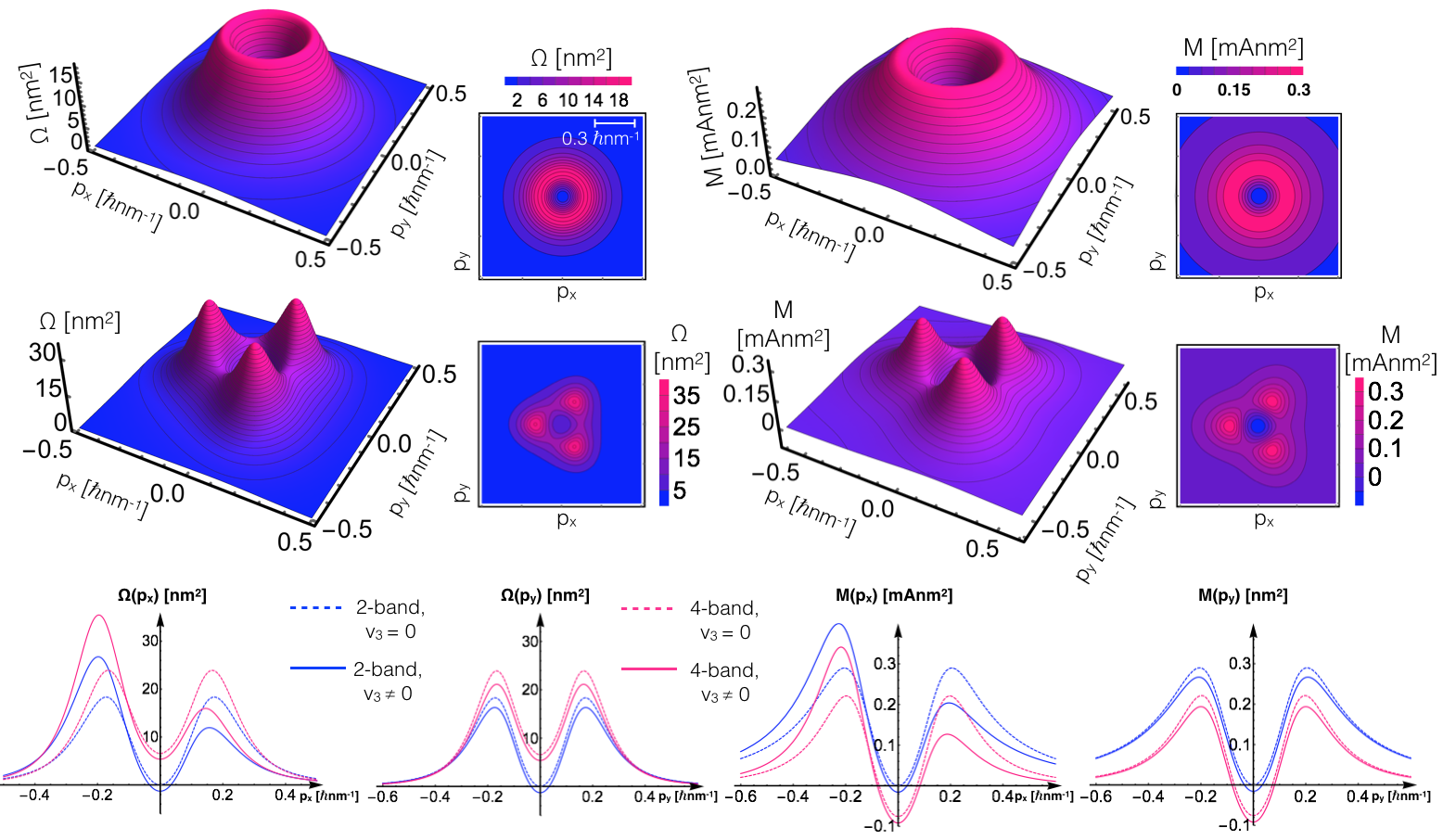}
 \caption{Berry curvature $\Omega$ and orbital magnetic moment $M$ of the lowest conduction band of homogeneous gapped \BLG calculated within various different levels of approximation. Top: $\Omega(p_x,p_y)$ and $M(p_x,p_y)$ within the two-band model without trigonal warping according to the analytical expressions of \eqn\ref{eqn:Berry}. Middle: $\Omega(p_x,p_y)$ and $M(p_x,p_y)$ within the four-band model including trigonal warping. Bottom: Cuts along the $p_x$, and $p_y$ direction, respectively, through $\Omega$ and $M$  for the different models.}
\label{fig:BerryS}
\end{figure*}

We discuss the topological properties of the homogeneous \BLG states we obtain from different models. In \fig\ref{fig:BerryS} we plot both magnitude of the Berry curvature, $\Omega$, and orbital magnetic moment, $M$, for the lowest conduction band of \BLG in the $K^-$ valley, computed for the two-band model or the four-band model with or without trigonal warping, respectively.
The two-band model description in the absence of trigonal warping reproduces the feature that Berry curvature and magnetic moment are non-zero only within a ring of finite width at finite momentum around the $K$-point. Within this approximation the magnetic moment is simply proportional to the Berry curvature (as for any particle-hole symmetry two-by-two Hamiltonian \cite{Xiao2007}). The anisotropic feature, however, that only the states in the minivalleys carry finite Berry curvature (and, consequently,  finite magnetic moment) are obtained only when $v_3$ is taken into account. The two-band model predicts the Berry curvature and the magnetic moment to vanish exactly at the graphene valley center, whereas in the  four-band model both quantities remain finite at zero momentum (but small as compared to the maximum value at the peaks). Within the two-band model in the absence of trigonal warping, $\Omega$ and $M$ can be computed analytically as
\begin{align}
\nonumber{\Omega}& \approx \xi\frac{\hbar^2}{2m}\frac{p^2}{2m}\frac{\Delta}{\epsilon(\mathbf{p})^3},\\
{M}& \approx  \epsilon(\mathbf{p}) \frac{e}{\hbar}\Omega=\xi\frac{e\hbar}{2m}\frac{p^2}{2m}\frac{\Delta}{\epsilon(\mathbf{p})^2}.
\label{eqn:Berry}
\end{align}
This allows to give an approximate formula for the maximal value of the magnetic moment which is simply given by
\begin{equation}
{M}_{max}\approx \mu =\xi\frac{e\hbar}{2m}.
\end{equation}
Hence we see that the minivalley states of gapped \BLG at zero magnetic field have finite Berry curvature and corresponding finite magnetic moment as shown in \fig\ref{fig:BerryS}. The orbital magnetic momentum behaves like the electron spin \cite{Xiao2010} and will therefore couple linearly to a magnetic field through a Zeeman-like term $-\mathbf{M}(\mathbf{k})\cdot\mathbf{B}$. This leads to the linear behaviour of the subbands in \fig\ref{fig:QPC_EwBTrigo} with $B$ with the same slope but opposite signs in the valleys $K^+$ and $K^-$.

\subsection{Adiabatic population of the transverse modes}
\label{sec:appPop}

\begin{figure}[h]
 \centering
\includegraphics[width=0.5\textwidth]{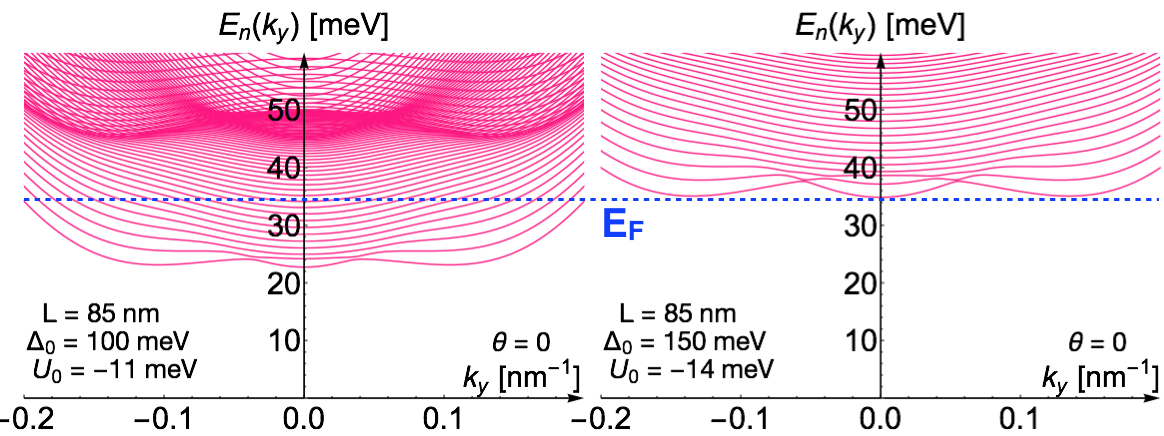}
 \caption{Adiabatic population of the channel modes with non-equilibrium electrons. When the channel depth is gradually increased, the modes of at the band edges of the channel spectra (Fermi energy $E_F$ in the right picture) become populated by non-equilibrium electrons that carry a non-zero transverse momentum (intersections with $E_F$ in the left picture).}
\label{fig:AdPop}
\end{figure}

We discuss the adiabatic population of the modes in the zero-field channel spectra by non-equilibrium electrons driven by a bias voltage applied at either end of the channel. This counting of transverse transport modes leads to the predictions for the conductance of the channel at zero magnetic field shown in \fig\ref{DispB0L50D60a250}. As an example we use the $\theta=0$ spectra for different $\Delta_0$ and $L$ shown in \fig\ref{fig:AdPop}. The electrons of homogeneous BLG (which would correspond to $U_0\equiv\Delta_0\equiv0$) are injected into the channel (which is realized for $U_0\neq0, \Delta_0\neq0$). Coupling in the modes adiabatically means increasing the confinement and the gap continuously. Figure \ref{fig:AdPop} demonstrates the evolution of the modes upon such continuous, gradual increase of the channel parameters: At the Fermi energy $E_F$ needed to populate the modes of the developing channel in the right hand side figure, in the left panel (corresponding to a more shallow part at the onset of the channel) only modes at non-zero momentum are available. Therefore, only the outer modes in the spectra (corresponding to the minivalleys B and C in \fig\ref{fig:Berry}) of the right hand side plot will be populated. Population of the zero-momentum modes would require relaxation of momentum via scattering processes. A similar line of argumentation also holds for the channel spectra for $\theta=\pi/1$, again leading to the conclusion that only the modes corresponding to the minivalleys at non-zero transverse momentum (minivalleys B and C) will be populated by non-equilibrium electrons.  Additional degeneracies between the the lowest energy levels, however, can lead to a higher number of modes available for the non-equilibrium electrons at the same energy. We observe such degeneracies for the bandedges of the lowest and the first conduction band for both angles above a certain critical value of $\Delta_0$. Due to this additional degeneracy we conjecture that there will be an additional factor of two for the conductance if $\Delta_0$ is large enough. 

\subsection{Perturbation Theory for small magnetic field strengths}
\label{sec:appPert}

As an additional consistency check,  the onset of the magnetic field effects for very small values of the magnetic field $B$ can be reproduced using perturbation theory. We  treat the magnetic part of the Hamiltonian as a perturbation writing $H^{\xi}_{BLG}=H^{\xi}_{0}+H^{\xi}_{B}$, where $H^{\xi}_{0}$ is the Hamiltonian in the absence of a magnetic field as given in \eqn (1), and the perturbation due to the magnetic field reads
\begin{align}
 H^{\xi}_{B}=
  \xi B\begin{pmatrix} 
0 &i\frac{e}{c} v_3 x & 0 &-i\frac{e}{c} v x\\
-i\frac{e}{c} v_3 x & 0 &i\frac{e}{c} v x &0\\
 0 &-i\frac{e}{c} v x&  0 &0\\
 i\frac{e}{c} v x & 0 & 0&0
\end{pmatrix}.
\label{eqn:HB}
 \end{align}

Due to the trigonal warping the lower band edges occur at a non-zero transverse momentum $(k_{x,min}, k_{y,min})$. Using these states at the band edges we compute the first order correction $ E_{1}$ to the energy  to read
\begin{align}
\nonumber &E^{\xi=+1}_{1}=\LD \Phi_{K^+}^0|  H^{\xi=+1}_{B} | \Phi_{K^+}^0  \RD \\
\nonumber=&2 \, B \frac{e}{c} \;  \mathfrak{Im} \Big[\, v\LD \Psi_{A }^0 |x|  \Psi_{B }^0\RD+  v\LD \Psi_{B^{\prime}}^0 |x|  \Psi_{A^{\prime} }^0\RD -v_3 \LD  \Psi_{A }^0 |x|  \Psi_{B^{\prime} }^0\RD \, \Big],\\
\nonumber &E^{\xi=-1}_{1}=\LD \Phi_{K^-}^0|  H^{\xi=-1}_{B} | \Phi_{K^-}^0  \RD \\
=&-2 \, B \frac{e}{c} \;  \mathfrak{Im} \Big[\, v\LD \Psi_{B^{\prime} }^0 |x|  \Psi_{A^{\prime} }^0\RD+  v\LD \Psi_{A}^0 |x|  \Psi_{B }^0\RD  -v_3 \LD  \Psi_{B^{\prime} }^0 |x|  \Psi_{A }^0\RD \, \Big],
\label{eqn:pertB}
\end{align}
where $| \Phi_{K^+/K^-}^0\RD$ denotes the unperturbed, valley degenerate state at zero magnetic field evaluated at finite $k_{y,min}$ as obtained from the numerics. It reproduces the gray, dashed lines in \fig 4. Hence the first order perturbation theory correction in the vector potential predicts the linear dependence on $B $ with opposite sign of the slope for either valley. This serves as an additional proof that it is the properties of the zero-field states in the minivalleys of gapped BLG (namely their non-trivial Berry curvature and finite magnetic moment) that dictate the sensitivity of the spectra to the magnetic field at low magnetic field (\textit{i.e.}, in the linear regime).

\subsection{Role of $\gamma_3$: Comparison of spectra without trigonal warping }
\label{sec:appTrigo}

\begin{figure}[h]
\includegraphics[width=0.3\textwidth]{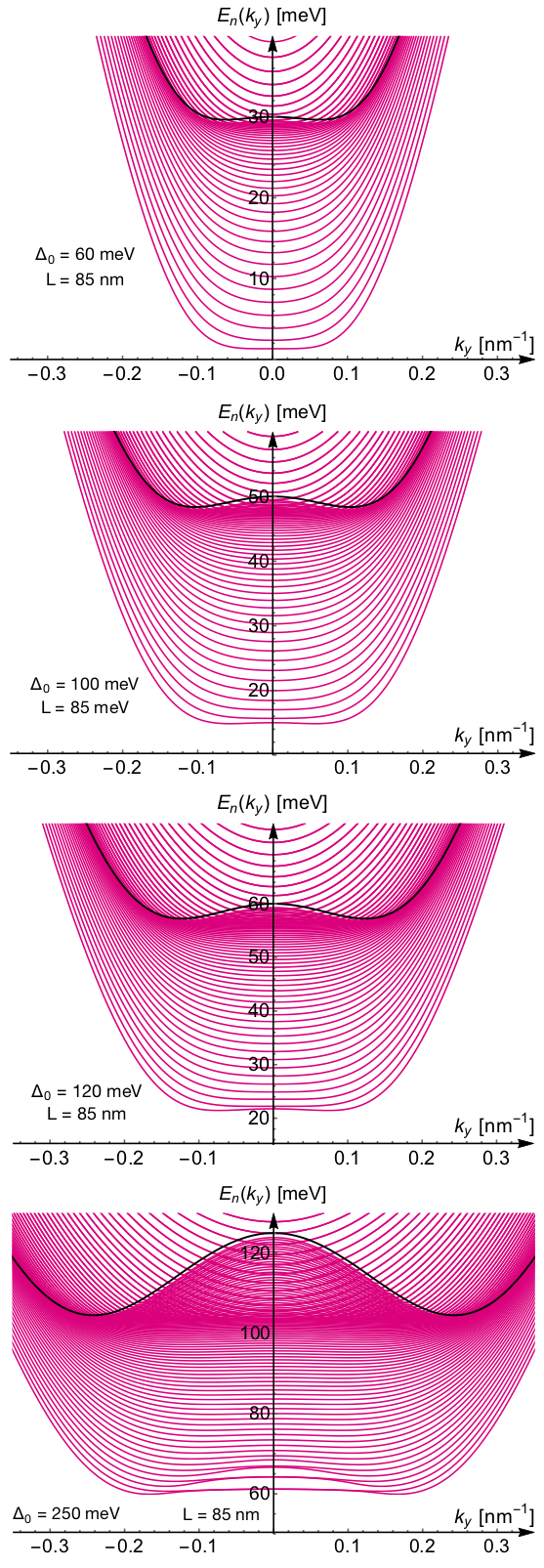}
 \caption{Energy levels of the conduction band for the \BLG \QPC without magnetic field for different values of   $\Delta_0$ when we do not consider trigonal warping effects and set $v_3\equiv0$. The black lines compare to the first conductance band of homogeneous gapped \BLG in the absence of confinement with a gap $\Delta_0$. }
\label{fig:QPC_SpectraNoBNT}
\end{figure} 

\begin{figure}[h!]
\includegraphics[width=0.3\textwidth]{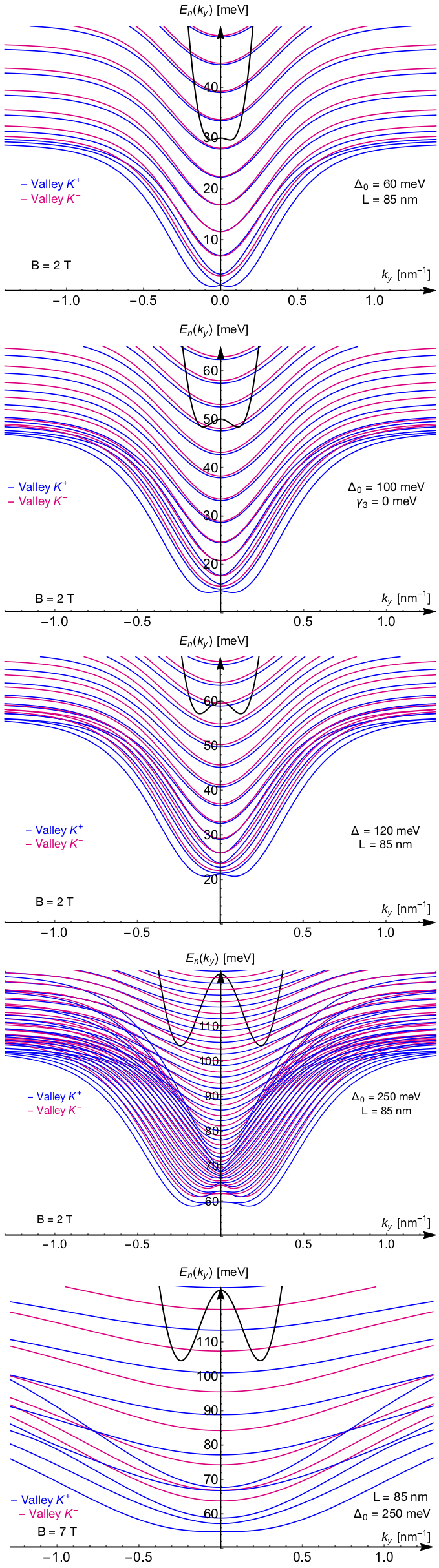}
 \caption{Landau levels (conduction band) of the \BLG \QPC for a magnetic field of $B=2$ T and different values of  $\Delta_0$ when we choose $v_3\equiv0$. Blue lines are for the $K^+$ valley, magenta lines are for the $K_-$ valley. The black lines compare to the first conductance band of homogeneous gapped \BLG in the absence of confinement with gap $\Delta_0$. }
\label{fig:QPC_SpectraWithBNT}
\end{figure}

\begin{figure}[h]
\includegraphics[width=0.4\textwidth]{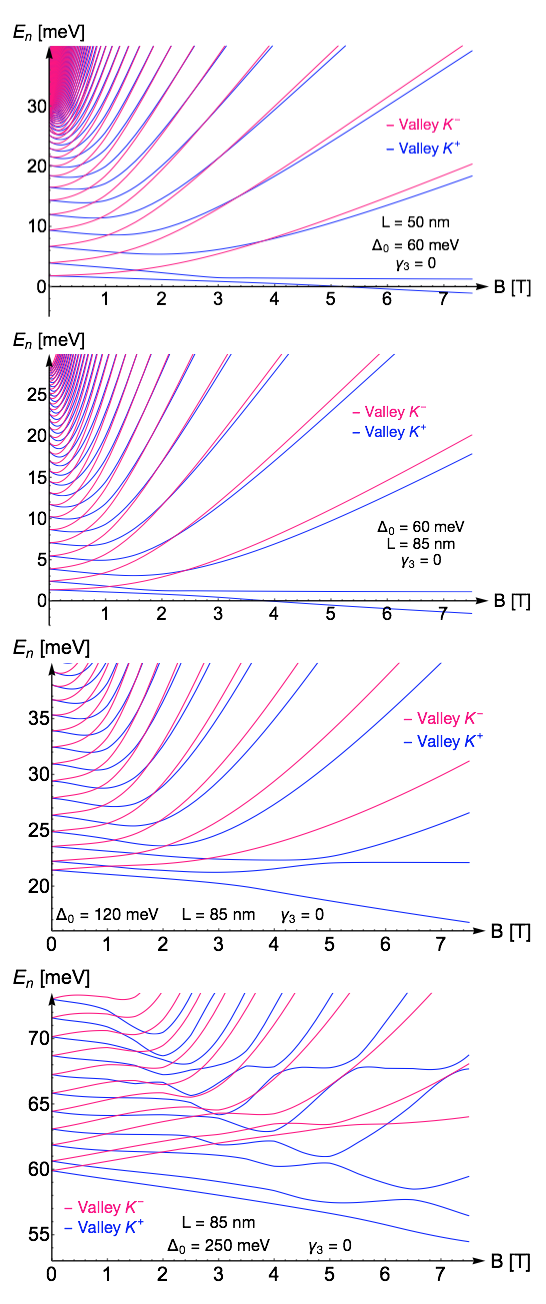}
 \caption{Conduction band edges as a function of the magnetic field for different  $L$ and different  values of  $\Delta_0$ for the case without trigonal warping effects, \textit{i.e.},  $v_3\equiv0$. Blue lines are for the valley $K^+$, magenta lines are for the valley $K^-$. }
\label{fig:QPC_EwB_AllNT}
\end{figure}

In \figs\ref{fig:QPC_SpectraNoBNT},  \ref{fig:QPC_SpectraWithBNT}, \ref{fig:QPC_EwB_AllNT}, we show the channel spectra with and without magnetic field in the absence of trigonal warping, \textit{i.e.}, when we chose $v_3\equiv0$, for a channel with $L=85$ nm and different values of $\Delta_0$ and the magnetic field. In the case without trigonal warping, the dispersion of homogeneous \BLG does not exhibit the threefold mini-valley structure, but is of rotationally symmetric Mexican-hat shape \cite{McCann2006}. Therefore, the channel spectra in this case do not depend on the angle of the channel orientation. We show in \figs \ref{fig:QPC_SpectraNoBNT} and \ref{fig:QPC_SpectraWithBNT} that the spectra do inherit the Mexican-hat features of the homogeneous gapped BLG's dispersion by developing a double minimum at non-zero $\pm k_{y,min}$. For smaller values of $\Delta_0$, the levels below the continuum are well separated and there are only  shallow modulations of the band edge which would be washed out by temperature fluctuations. Hence, effectively, a single, non-degenerate minimum is formed. For larger $\Delta_0$, the modes become degenerate at $k_y=0$, but are pushed apart  at non-zero momentum, in analogy to tunneling processes and instantons in double-well potentials. As a consequence, a clearly separated double minimum develops at non-zero value of the transverse momentum. As discussed earlier, at non-zero momentum the states of \BLG have non-zero Berry curvature and non-zero orbital magnetic moment, even for $v_3=0$. As a consequence, also in the absence of trigonal warping, we see linear valley splitting with magnetic field in small magnetic fields, see \fig\ref{fig:QPC_EwB_AllNT}.

\begin{figure}[h!]
 \centering
\includegraphics[width=0.45\textwidth]{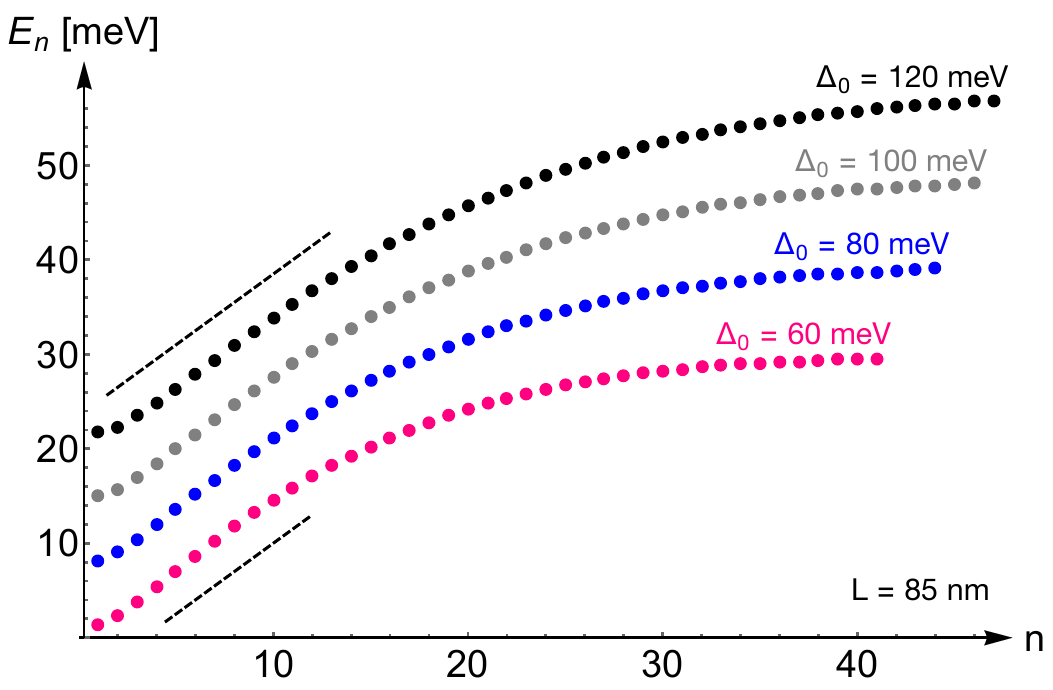}
 \caption{Properties of the spectra at zero magnetic field. The lowest energy levels $E_{n\lesssim12}$ scale linearly with the level number $n$, where the slope is independent of $\Delta_0$.  }
\label{fig:QPC_SpectraStats}
\end{figure}

 
 We demonstrate some of the properties of the zero-field levels in  \fig\ref{fig:QPC_SpectraStats} where we show that the lowest energy levels $E_{n\lesssim12}$ follow an approximately linear dependence on the level number $n$, where the slope is independent of the gap. The level spacing in this regime is found  to scale like $1/L$ as a reminiscence of the near-to-quadratic confinement. 
 


\subsection{Effective masses of the BLG minivalleys}
\label{sec:appMass}

\begin{figure}[h]
 \centering
\includegraphics[width=0.5\textwidth]{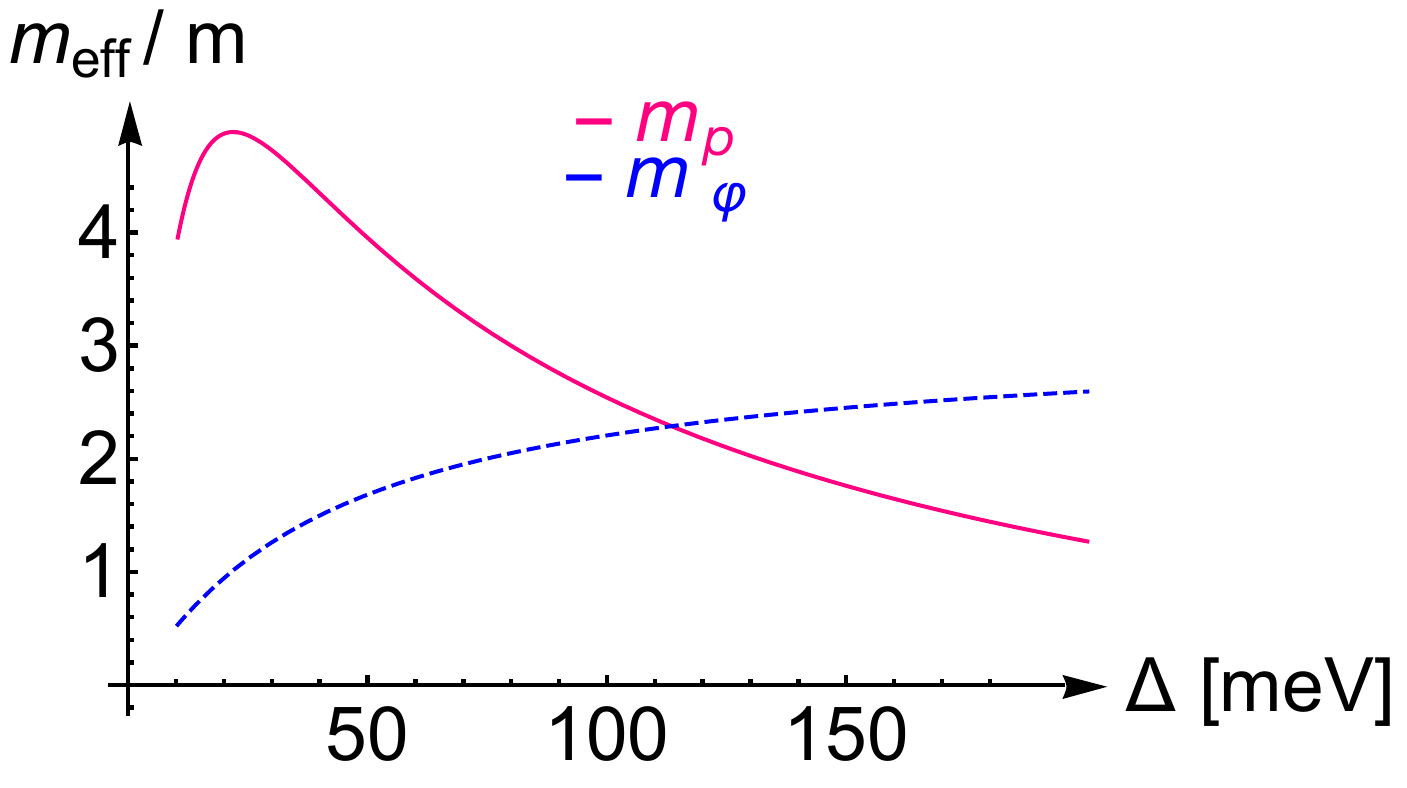}
 \caption{Effective radial and azimuthal masses of the minivalleys in the \BLG dispersion as a function of $\Delta$ (in relation to the mass $m$ of electron in gapless, BLG).}
\label{fig:QPC_massplot}
\end{figure}

Within the two-band model approximation the dispersion of gapped \BLG reads \cite{Li2014}
\begin{align}
\nonumber E=\pm& \Bigg(\frac{{\Delta}^2}{4}+\frac{\left({\delta}^2+1\right) p^4 v^4}{{\gamma_1}^2}+p^2 \left(v^2 {\tau}^2-{\delta}^2 v^2\right)\\
-&\frac{2 p^3 v^3 {\tau} \cos (3 {\varphi})}{{\gamma_1}} \Bigg)^{\frac{1}{2}},
\end{align}
where $\delta=\frac{\Delta}{\gamma_1}$. We consider the limit of small, but finite gaps ($\delta\ll1$, but $\Delta>v_3^2m$) and small $\tau$ ($\tau<<1$). The conduction band assumes three minima at radius 
\begin{align}
\nonumber p_0&=\frac{{\gamma_1} \left(\sqrt{8 {\delta}^2 \left({\delta}^2-{\tau}^2+1\right)+{\tau}^2}+3 {\tau}\right)}{4 \left({\delta}^2+1\right) v} \\
&\approx \frac{\gamma_1}{4v}[\sqrt{8\delta^2+\tau^2}+3\tau],
\end{align}
 and angles $\varphi=0, \frac{2}{3}\pi, \frac{4}{3}\pi$ (in the $K^+$ valley) or $\varphi=\pi, \frac{1}{3}\pi, \frac{5}{3}\pi$ (in the $K^-$ valley). For small variations around the minima, an expansion to second order in $p$ and $\varphi$ around $p_0$ yields the approximate expansion coefficients 
 \begin{align}
\nonumber a_{\varphi}&\approx \frac{9}{64}\frac{\gamma_1^2 \tau}{\Delta}[\sqrt{8\delta^2+\tau^2}+3\tau]^3,\\
 a_{p}&\approx \frac{1}{4}\frac{v^2}{\Delta}[8\delta^4+8\delta^2+\tau(3\sqrt{8\delta^2+\tau^2}+\tau)],
\end{align}
from which the approximate effective azimuthal and radial masses are computed
\begin{align}
 m_{\varphi}&\approx \frac{4}{9}\frac{\Delta}{v^2} \frac{1}{ \tau(\sqrt{8D^2+\tau^2}+3\tau)},\\
  m_{p}& \approx 4\frac{\Delta}{v^2}\frac{1}{8\delta^4+8\delta^2+\tau(3\sqrt{8\delta^2+\tau^2}+\tau)}.
\end{align}
Figure \ref{fig:QPC_massplot} shows both effective masses as a function of $\Delta$.

 \section{Conclusion}
In summary,
we find that the minivalley structure of the gapped \BLG dispersion determines the properties of electrostatically defined 1D transport channels in devices based on this 2D material. The anisotropy of the dispersion in and around the minivalleys makes the subband spectra sensitive to the orientation of the channel, including the position and the splitting of the band edges, therefore determining the sequence of quantized conductance steps in the BLG wire. 

Depending on the channel width and the chrystallographic orientation, and the size of the gap in BLG, we find that the first step developing upon filling the channel with electrons may have $8e^2/h$ or $4e^2/h$ height, for larger or smaller gaps $\Delta$, respectively. Also, we find that the miniband splitting linear in magnetic field, prescribed by the finite magnetic moment of the electron states at the minivalleys of gapped BLG (related to the Berry curvature of the gapped BLG bands), results in a conductance staircase of a sequence of $2e^2/h$ steps.

While the above results identify the generic degeneracies in the subbands spectra of electrostatically defined wires in gapped BLG at the singe-particle level (which would not be affected by self-consistent Hartree effects produced by electron-electron interaction on the shape of voltage potential in the channel), the details of the potential profile, $U(x)$ and the gap modulation, $\Delta(x)$, would affect the numerical values of the subbands offset energies. Also, the exchange interaction in gapped BLG is known to increase the single-particle band inversion near the centre of BLG valleys \cite{Cheianov1012}, making minivalleys more pronounced spectral features in a homogeneously gapped BLG, so that we expect that the minivalley effects predicted by the single-particle theory would be more pronounced in the real experimentally studied systems, at least at the lowest temperatures. 
Moreover, one may expect that electron-electron interaction in the weakly filled lowest subband of a BLG quantum wire may produce effects similar to the thermally activated 0.7 anomaly in GaAs quantum wires \cite{Thomas1996, Thomas1998, Micolich2011, Hamilton2008}. However, the very flat and, in some parametric ranges, non-monotonic (even slightly inverted around $k_y=0$) lowest energy subbands shown in Figs. 3-10 point towards other possible roles that may be played  by electron-electron interaction. For one, low-density electrons in a long BLG wire may form a charge-density wave ground state (1D Wigner crystals). Another option would be that, for a partly filled lowest subband, its slightly inverted dispersion with a maximum at $k_y=0$ (Figs. 3, 5, and 10) may open new electron-electron scattering channels between non-equilbrium electrons injected from bulk electrodes, as discussed in subsection III.C, and electrons at the Fermi energy in the 'passive' central part of the subband dispersion, which would  lead to a strong suppression of the quantum wire conductance at intermediate temperatures. These interesting possible effects of electron-electron interaction on the BLG quantum wire characteristics will be subject of further studies.

\begin{acknowledgments}
We would like to thank H. Overweg, M. Eich, K. Ensslin,  T. Ihn, E. McCann, X. Chen, S. Slizovskiy, J. R. Wallbank, D. Ruiz-Tijerina, and T. L. M. Lane for discussions. We acknowledge funding from the ERC Synergy Grant, the European Quantum Technology Flagship,  and the European Graphene Flagship.
\end{acknowledgments}


\bibliography{QPC}

\begin{thebibliography}{36}%
\makeatletter
\providecommand \@ifxundefined [1]{%
 \@ifx{#1\undefined}
}%
\providecommand \@ifnum [1]{%
 \ifnum #1\expandafter \@firstoftwo
 \else \expandafter \@secondoftwo
 \fi
}%
\providecommand \@ifx [1]{%
 \ifx #1\expandafter \@firstoftwo
 \else \expandafter \@secondoftwo
 \fi
}%
\providecommand \natexlab [1]{#1}%
\providecommand \enquote  [1]{``#1''}%
\providecommand \bibnamefont  [1]{#1}%
\providecommand \bibfnamefont [1]{#1}%
\providecommand \citenamefont [1]{#1}%
\providecommand \href@noop [0]{\@secondoftwo}%
\providecommand \href [0]{\begingroup \@sanitize@url \@href}%
\providecommand \@href[1]{\@@startlink{#1}\@@href}%
\providecommand \@@href[1]{\endgroup#1\@@endlink}%
\providecommand \@sanitize@url [0]{\catcode `\\12\catcode `\$12\catcode
  `\&12\catcode `\#12\catcode `\^12\catcode `\_12\catcode `\%12\relax}%
\providecommand \@@startlink[1]{}%
\providecommand \@@endlink[0]{}%
\providecommand \url  [0]{\begingroup\@sanitize@url \@url }%
\providecommand \@url [1]{\endgroup\@href {#1}{\urlprefix }}%
\providecommand \urlprefix  [0]{URL }%
\providecommand \Eprint [0]{\href }%
\providecommand \doibase [0]{http://dx.doi.org/}%
\providecommand \selectlanguage [0]{\@gobble}%
\providecommand \bibinfo  [0]{\@secondoftwo}%
\providecommand \bibfield  [0]{\@secondoftwo}%
\providecommand \translation [1]{[#1]}%
\providecommand \BibitemOpen [0]{}%
\providecommand \bibitemStop [0]{}%
\providecommand \bibitemNoStop [0]{.\EOS\space}%
\providecommand \EOS [0]{\spacefactor3000\relax}%
\providecommand \BibitemShut  [1]{\csname bibitem#1\endcsname}%
\let\auto@bib@innerbib\@empty
\bibitem [{\citenamefont {Loss}\ and\ \citenamefont
  {DiVincenzo}(1998)}]{Loss1998}%
  \BibitemOpen
  \bibfield  {author} {\bibinfo {author} {\bibfnamefont {D.}~\bibnamefont
  {Loss}}\ and\ \bibinfo {author} {\bibfnamefont {D.~P.}\ \bibnamefont
  {DiVincenzo}},\ }\href {\doibase 10.1103/PhysRevA.57.120} {\bibfield
  {journal} {\bibinfo  {journal} {Physical Review A}\ }\textbf {\bibinfo
  {volume} {57}},\ \bibinfo {pages} {120} (\bibinfo {year} {1998})}\BibitemShut
  {NoStop}%
\bibitem [{\citenamefont {Fischer}\ \emph {et~al.}(2009)\citenamefont
  {Fischer}, \citenamefont {Trauzettel},\ and\ \citenamefont
  {Loss}}]{Fischer2009}%
  \BibitemOpen
  \bibfield  {author} {\bibinfo {author} {\bibfnamefont {J.}~\bibnamefont
  {Fischer}}, \bibinfo {author} {\bibfnamefont {B.}~\bibnamefont {Trauzettel}},
  \ and\ \bibinfo {author} {\bibfnamefont {D.}~\bibnamefont {Loss}},\ }\href
  {\doibase 10.1103/PhysRevB.80.155401} {\bibfield  {journal} {\bibinfo
  {journal} {Physical Review B}\ }\textbf {\bibinfo {volume} {80}},\ \bibinfo
  {pages} {155401} (\bibinfo {year} {2009})}\BibitemShut {NoStop}%
\bibitem [{\citenamefont {Yao}\ \emph {et~al.}(2006)\citenamefont {Yao},
  \citenamefont {Liu},\ and\ \citenamefont {Sham}}]{Yao2006}%
  \BibitemOpen
  \bibfield  {author} {\bibinfo {author} {\bibfnamefont {W.}~\bibnamefont
  {Yao}}, \bibinfo {author} {\bibfnamefont {R.-B.}\ \bibnamefont {Liu}}, \ and\
  \bibinfo {author} {\bibfnamefont {L.~J.}\ \bibnamefont {Sham}},\ }\href
  {\doibase 10.1103/PhysRevB.74.195301} {\bibfield  {journal} {\bibinfo
  {journal} {Physical Review B}\ }\textbf {\bibinfo {volume} {74}},\ \bibinfo
  {pages} {195301} (\bibinfo {year} {2006})}\BibitemShut {NoStop}%
\bibitem [{\citenamefont {Klauser}\ \emph {et~al.}(2007)\citenamefont
  {Klauser}, \citenamefont {Bulaev}, \citenamefont {Coish},\ and\ \citenamefont
  {Loss}}]{Klauser2007}%
  \BibitemOpen
  \bibfield  {author} {\bibinfo {author} {\bibfnamefont {D.}~\bibnamefont
  {Klauser}}, \bibinfo {author} {\bibfnamefont {D.~V.}\ \bibnamefont {Bulaev}},
  \bibinfo {author} {\bibfnamefont {W.~A.}\ \bibnamefont {Coish}}, \ and\
  \bibinfo {author} {\bibfnamefont {D.}~\bibnamefont {Loss}},\ }\href@noop {}
  {\bibfield  {journal} {\bibinfo  {journal} {arXiv:0706.1514 [cond-mat]}\ }
  (\bibinfo {year} {2007})},\ \Eprint {http://arxiv.org/abs/0706.1514}
  {arXiv:0706.1514 [cond-mat]} \BibitemShut {NoStop}%
\bibitem [{\citenamefont {Rycerz}\ \emph {et~al.}(2007)\citenamefont {Rycerz},
  \citenamefont {Tworzyd\l{}o},\ and\ \citenamefont {Beenakker}}]{Rycerz2007}%
  \BibitemOpen
  \bibfield  {author} {\bibinfo {author} {\bibfnamefont {A.}~\bibnamefont
  {Rycerz}}, \bibinfo {author} {\bibfnamefont {J.}~\bibnamefont
  {Tworzyd\l{}o}}, \ and\ \bibinfo {author} {\bibfnamefont {C.~W.~J.}\
  \bibnamefont {Beenakker}},\ }\href@noop {} {\bibfield  {journal} {\bibinfo
  {journal} {Nature Physics}\ }\textbf {\bibinfo {volume} {3}},\ \bibinfo
  {pages} {172} (\bibinfo {year} {2007})}\BibitemShut {NoStop}%
\bibitem [{\citenamefont {Trauzettel}\ \emph {et~al.}(2007)\citenamefont
  {Trauzettel}, \citenamefont {Bulaev}, \citenamefont {Loss},\ and\
  \citenamefont {Burkard}}]{Trauzettel2007}%
  \BibitemOpen
  \bibfield  {author} {\bibinfo {author} {\bibfnamefont {B.}~\bibnamefont
  {Trauzettel}}, \bibinfo {author} {\bibfnamefont {D.~V.}\ \bibnamefont
  {Bulaev}}, \bibinfo {author} {\bibfnamefont {D.}~\bibnamefont {Loss}}, \ and\
  \bibinfo {author} {\bibfnamefont {G.}~\bibnamefont {Burkard}},\ }\href@noop
  {} {\bibfield  {journal} {\bibinfo  {journal} {Nature Physics}\ }\textbf
  {\bibinfo {volume} {3}},\ \bibinfo {pages} {192} (\bibinfo {year}
  {2007})}\BibitemShut {NoStop}%
\bibitem [{\citenamefont {{Bischoff Dominik}}\ \emph
  {et~al.}(2015)\citenamefont {{Bischoff Dominik}}, \citenamefont {{Simonet
  Pauline}}, \citenamefont {{Varlet Anastasia}}, \citenamefont {{Overweg Hiske
  C.}}, \citenamefont {{Eich Marius}}, \citenamefont {{Ihn Thomas}},\ and\
  \citenamefont {{Ensslin Klaus}}}]{BischoffDominik2015}%
  \BibitemOpen
  \bibfield  {author} {\bibinfo {author} {\bibnamefont {{Bischoff Dominik}}},
  \bibinfo {author} {\bibnamefont {{Simonet Pauline}}}, \bibinfo {author}
  {\bibnamefont {{Varlet Anastasia}}}, \bibinfo {author} {\bibnamefont
  {{Overweg Hiske C.}}}, \bibinfo {author} {\bibnamefont {{Eich Marius}}},
  \bibinfo {author} {\bibnamefont {{Ihn Thomas}}}, \ and\ \bibinfo {author}
  {\bibnamefont {{Ensslin Klaus}}},\ }\href {\doibase 10.1002/pssr.201510163}
  {\bibfield  {journal} {\bibinfo  {journal} {physica status solidi (RRL)
  \textendash{} Rapid Research Letters}\ }\textbf {\bibinfo {volume} {10}},\
  \bibinfo {pages} {68} (\bibinfo {year} {2015})}\BibitemShut {NoStop}%
\bibitem [{\citenamefont {McCann}\ and\ \citenamefont
  {Fal'ko}(2006)}]{McCann2006}%
  \BibitemOpen
  \bibfield  {author} {\bibinfo {author} {\bibfnamefont {E.}~\bibnamefont
  {McCann}}\ and\ \bibinfo {author} {\bibfnamefont {V.~I.}\ \bibnamefont
  {Fal'ko}},\ }\href {\doibase 10.1103/PhysRevLett.96.086805} {\bibfield
  {journal} {\bibinfo  {journal} {Physical Review Letters}\ }\textbf {\bibinfo
  {volume} {96}},\ \bibinfo {pages} {086805} (\bibinfo {year}
  {2006})}\BibitemShut {NoStop}%
\bibitem [{\citenamefont {Heersche}\ \emph {et~al.}(2007)\citenamefont
  {Heersche}, \citenamefont {{Jarillo-Herrero}}, \citenamefont {Oostinga},
  \citenamefont {Vandersypen},\ and\ \citenamefont {Morpurgo}}]{Heersche2007}%
  \BibitemOpen
  \bibfield  {author} {\bibinfo {author} {\bibfnamefont {H.~B.}\ \bibnamefont
  {Heersche}}, \bibinfo {author} {\bibfnamefont {P.}~\bibnamefont
  {{Jarillo-Herrero}}}, \bibinfo {author} {\bibfnamefont {J.~B.}\ \bibnamefont
  {Oostinga}}, \bibinfo {author} {\bibfnamefont {L.~M.~K.}\ \bibnamefont
  {Vandersypen}}, \ and\ \bibinfo {author} {\bibfnamefont {A.~F.}\ \bibnamefont
  {Morpurgo}},\ }\href@noop {} {\bibfield  {journal} {\bibinfo  {journal}
  {Nature}\ }\textbf {\bibinfo {volume} {446}},\ \bibinfo {pages} {56}
  (\bibinfo {year} {2007})}\BibitemShut {NoStop}%
\bibitem [{\citenamefont {Varlet}\ \emph {et~al.}(2014)\citenamefont {Varlet},
  \citenamefont {Bischoff}, \citenamefont {Simonet}, \citenamefont {Watanabe},
  \citenamefont {Taniguchi}, \citenamefont {Ihn}, \citenamefont {Ensslin},
  \citenamefont {{Mucha-Kruczy\'nski}},\ and\ \citenamefont
  {Fal'ko}}]{Varlet2014}%
  \BibitemOpen
  \bibfield  {author} {\bibinfo {author} {\bibfnamefont {A.}~\bibnamefont
  {Varlet}}, \bibinfo {author} {\bibfnamefont {D.}~\bibnamefont {Bischoff}},
  \bibinfo {author} {\bibfnamefont {P.}~\bibnamefont {Simonet}}, \bibinfo
  {author} {\bibfnamefont {K.}~\bibnamefont {Watanabe}}, \bibinfo {author}
  {\bibfnamefont {T.}~\bibnamefont {Taniguchi}}, \bibinfo {author}
  {\bibfnamefont {T.}~\bibnamefont {Ihn}}, \bibinfo {author} {\bibfnamefont
  {K.}~\bibnamefont {Ensslin}}, \bibinfo {author} {\bibfnamefont
  {M.}~\bibnamefont {{Mucha-Kruczy\'nski}}}, \ and\ \bibinfo {author}
  {\bibfnamefont {V.~I.}\ \bibnamefont {Fal'ko}},\ }\href {\doibase
  10.1103/PhysRevLett.113.116602} {\bibfield  {journal} {\bibinfo  {journal}
  {Physical Review Letters}\ }\textbf {\bibinfo {volume} {113}},\ \bibinfo
  {pages} {116602} (\bibinfo {year} {2014})}\BibitemShut {NoStop}%
\bibitem [{\citenamefont {Varlet}\ \emph {et~al.}(2015)\citenamefont {Varlet},
  \citenamefont {{Mucha-Kruczy\'nski}}, \citenamefont {Bischoff}, \citenamefont
  {Simonet}, \citenamefont {Taniguchi}, \citenamefont {Watanabe}, \citenamefont
  {Fal'ko}, \citenamefont {Ihn},\ and\ \citenamefont {Ensslin}}]{Varlet2015}%
  \BibitemOpen
  \bibfield  {author} {\bibinfo {author} {\bibfnamefont {A.}~\bibnamefont
  {Varlet}}, \bibinfo {author} {\bibfnamefont {M.}~\bibnamefont
  {{Mucha-Kruczy\'nski}}}, \bibinfo {author} {\bibfnamefont {D.}~\bibnamefont
  {Bischoff}}, \bibinfo {author} {\bibfnamefont {P.}~\bibnamefont {Simonet}},
  \bibinfo {author} {\bibfnamefont {T.}~\bibnamefont {Taniguchi}}, \bibinfo
  {author} {\bibfnamefont {K.}~\bibnamefont {Watanabe}}, \bibinfo {author}
  {\bibfnamefont {V.}~\bibnamefont {Fal'ko}}, \bibinfo {author} {\bibfnamefont
  {T.}~\bibnamefont {Ihn}}, \ and\ \bibinfo {author} {\bibfnamefont
  {K.}~\bibnamefont {Ensslin}},\ }\href {\doibase
  10.1016/j.synthmet.2015.07.006} {\bibfield  {journal} {\bibinfo  {journal}
  {Synthetic Metals}\ }\bibinfo {series} {Reviews of Current Advances in
  Graphene Science and Technology},\ \textbf {\bibinfo {volume} {210}},\
  \bibinfo {pages} {19} (\bibinfo {year} {2015})}\BibitemShut {NoStop}%
\bibitem [{\citenamefont {Fal'ko}(2007)}]{Falko2007}%
  \BibitemOpen
  \bibfield  {author} {\bibinfo {author} {\bibfnamefont {V.}~\bibnamefont
  {Fal'ko}},\ }\href@noop {} {\bibfield  {journal} {\bibinfo  {journal} {Nature
  Physics}\ }\textbf {\bibinfo {volume} {3}},\ \bibinfo {pages} {151} (\bibinfo
  {year} {2007})}\BibitemShut {NoStop}%
\bibitem [{\citenamefont {Dr\"oscher}\ \emph {et~al.}(2012)\citenamefont
  {Dr\"oscher}, \citenamefont {Barraud}, \citenamefont {Watanabe},
  \citenamefont {Taniguchi}, \citenamefont {Ihn},\ and\ \citenamefont
  {Ensslin}}]{Droscher2012}%
  \BibitemOpen
  \bibfield  {author} {\bibinfo {author} {\bibfnamefont {S.}~\bibnamefont
  {Dr\"oscher}}, \bibinfo {author} {\bibfnamefont {C.}~\bibnamefont {Barraud}},
  \bibinfo {author} {\bibfnamefont {K.}~\bibnamefont {Watanabe}}, \bibinfo
  {author} {\bibfnamefont {T.}~\bibnamefont {Taniguchi}}, \bibinfo {author}
  {\bibfnamefont {T.}~\bibnamefont {Ihn}}, \ and\ \bibinfo {author}
  {\bibfnamefont {K.}~\bibnamefont {Ensslin}},\ }\href {\doibase
  10.1088/1367-2630/14/10/103007} {\bibfield  {journal} {\bibinfo  {journal}
  {New Journal of Physics}\ }\textbf {\bibinfo {volume} {14}},\ \bibinfo
  {pages} {103007} (\bibinfo {year} {2012})}\BibitemShut {NoStop}%
\bibitem [{\citenamefont {Overweg}\ \emph {et~al.}(2018)\citenamefont
  {Overweg}, \citenamefont {Eggimann}, \citenamefont {Chen}, \citenamefont
  {Slizovskiy}, \citenamefont {Eich}, \citenamefont {Pisoni}, \citenamefont
  {Lee}, \citenamefont {Rickhaus}, \citenamefont {Watanabe}, \citenamefont
  {Taniguchi}, \citenamefont {Fal'ko}, \citenamefont {Ihn},\ and\ \citenamefont
  {Ensslin}}]{Overweg2018}%
  \BibitemOpen
  \bibfield  {author} {\bibinfo {author} {\bibfnamefont {H.}~\bibnamefont
  {Overweg}}, \bibinfo {author} {\bibfnamefont {H.}~\bibnamefont {Eggimann}},
  \bibinfo {author} {\bibfnamefont {X.}~\bibnamefont {Chen}}, \bibinfo {author}
  {\bibfnamefont {S.}~\bibnamefont {Slizovskiy}}, \bibinfo {author}
  {\bibfnamefont {M.}~\bibnamefont {Eich}}, \bibinfo {author} {\bibfnamefont
  {R.}~\bibnamefont {Pisoni}}, \bibinfo {author} {\bibfnamefont
  {Y.}~\bibnamefont {Lee}}, \bibinfo {author} {\bibfnamefont {P.}~\bibnamefont
  {Rickhaus}}, \bibinfo {author} {\bibfnamefont {K.}~\bibnamefont {Watanabe}},
  \bibinfo {author} {\bibfnamefont {T.}~\bibnamefont {Taniguchi}}, \bibinfo
  {author} {\bibfnamefont {V.}~\bibnamefont {Fal'ko}}, \bibinfo {author}
  {\bibfnamefont {T.}~\bibnamefont {Ihn}}, \ and\ \bibinfo {author}
  {\bibfnamefont {K.}~\bibnamefont {Ensslin}},\ }\href {\doibase
  10.1021/acs.nanolett.7b04666} {\bibfield  {journal} {\bibinfo  {journal}
  {Nano Letters}\ }\textbf {\bibinfo {volume} {18}},\ \bibinfo {pages} {553}
  (\bibinfo {year} {2018})}\BibitemShut {NoStop}%
\bibitem [{\citenamefont {Kraft}\ \emph {et~al.}(2018)\citenamefont {Kraft},
  \citenamefont {Krainov}, \citenamefont {V.~Gall}, \citenamefont {Dmitriev},
  \citenamefont {Krupke}, \citenamefont {Gornyi},\ and\ \citenamefont
  {Danneau}}]{Kraft2018}%
  \BibitemOpen
  \bibfield  {author} {\bibinfo {author} {\bibfnamefont {R.}~\bibnamefont
  {Kraft}}, \bibinfo {author} {\bibfnamefont {I.}~\bibnamefont {Krainov}},
  \bibinfo {author} {\bibfnamefont {V.}~\bibnamefont {V.~Gall}}, \bibinfo
  {author} {\bibfnamefont {A.}~\bibnamefont {Dmitriev}}, \bibinfo {author}
  {\bibfnamefont {R.}~\bibnamefont {Krupke}}, \bibinfo {author} {\bibfnamefont
  {I.}~\bibnamefont {Gornyi}}, \ and\ \bibinfo {author} {\bibfnamefont
  {R.}~\bibnamefont {Danneau}},\ }\href@noop {} {\bibfield  {journal} {\bibinfo
   {journal} {arXiv:1809.02458 [cond-mat]}\ } (\bibinfo {year} {2018})},\
  \Eprint {http://arxiv.org/abs/1809.02458} {arXiv:1809.02458 [cond-mat]}
  \BibitemShut {NoStop}%
\bibitem [{\citenamefont {Hunt}\ \emph {et~al.}(2017)\citenamefont {Hunt},
  \citenamefont {Li}, \citenamefont {Zibrov}, \citenamefont {Wang},
  \citenamefont {Taniguchi}, \citenamefont {Watanabe}, \citenamefont {Hone},
  \citenamefont {Dean}, \citenamefont {Zaletel}, \citenamefont {Ashoori},\ and\
  \citenamefont {Young}}]{Hunt2017}%
  \BibitemOpen
  \bibfield  {author} {\bibinfo {author} {\bibfnamefont {B.~M.}\ \bibnamefont
  {Hunt}}, \bibinfo {author} {\bibfnamefont {J.~I.~A.}\ \bibnamefont {Li}},
  \bibinfo {author} {\bibfnamefont {A.~A.}\ \bibnamefont {Zibrov}}, \bibinfo
  {author} {\bibfnamefont {L.}~\bibnamefont {Wang}}, \bibinfo {author}
  {\bibfnamefont {T.}~\bibnamefont {Taniguchi}}, \bibinfo {author}
  {\bibfnamefont {K.}~\bibnamefont {Watanabe}}, \bibinfo {author}
  {\bibfnamefont {J.}~\bibnamefont {Hone}}, \bibinfo {author} {\bibfnamefont
  {C.~R.}\ \bibnamefont {Dean}}, \bibinfo {author} {\bibfnamefont
  {M.}~\bibnamefont {Zaletel}}, \bibinfo {author} {\bibfnamefont {R.~C.}\
  \bibnamefont {Ashoori}}, \ and\ \bibinfo {author} {\bibfnamefont {A.~F.}\
  \bibnamefont {Young}},\ }\href {\doibase 10.1038/s41467-017-00824-w}
  {\bibfield  {journal} {\bibinfo  {journal} {Nature Communications}\ }\textbf
  {\bibinfo {volume} {8}},\ \bibinfo {pages} {948} (\bibinfo {year}
  {2017})}\BibitemShut {NoStop}%
\bibitem [{\citenamefont {Allen}\ \emph {et~al.}(2012)\citenamefont {Allen},
  \citenamefont {Martin},\ and\ \citenamefont {Yacoby}}]{Allen2012}%
  \BibitemOpen
  \bibfield  {author} {\bibinfo {author} {\bibfnamefont {M.~T.}\ \bibnamefont
  {Allen}}, \bibinfo {author} {\bibfnamefont {J.}~\bibnamefont {Martin}}, \
  and\ \bibinfo {author} {\bibfnamefont {A.}~\bibnamefont {Yacoby}},\
  }\href@noop {} {\bibfield  {journal} {\bibinfo  {journal} {Nature
  Communications}\ }\textbf {\bibinfo {volume} {3}},\ \bibinfo {pages} {934}
  (\bibinfo {year} {2012})}\BibitemShut {NoStop}%
\bibitem [{\citenamefont {Goossens}\ \emph {et~al.}(2012)\citenamefont
  {Goossens}, \citenamefont {Driessen}, \citenamefont {Baart}, \citenamefont
  {Watanabe}, \citenamefont {Taniguchi},\ and\ \citenamefont
  {Vandersypen}}]{Goossens2012}%
  \BibitemOpen
  \bibfield  {author} {\bibinfo {author} {\bibfnamefont {A.~S.~M.}\
  \bibnamefont {Goossens}}, \bibinfo {author} {\bibfnamefont {S.~C.~M.}\
  \bibnamefont {Driessen}}, \bibinfo {author} {\bibfnamefont {T.~A.}\
  \bibnamefont {Baart}}, \bibinfo {author} {\bibfnamefont {K.}~\bibnamefont
  {Watanabe}}, \bibinfo {author} {\bibfnamefont {T.}~\bibnamefont {Taniguchi}},
  \ and\ \bibinfo {author} {\bibfnamefont {L.~M.~K.}\ \bibnamefont
  {Vandersypen}},\ }\href {\doibase 10.1021/nl301986q} {\bibfield  {journal}
  {\bibinfo  {journal} {Nano Letters}\ }\textbf {\bibinfo {volume} {12}},\
  \bibinfo {pages} {4656} (\bibinfo {year} {2012})}\BibitemShut {NoStop}%
\bibitem [{\citenamefont {Eich}\ \emph {et~al.}(2018)\citenamefont {Eich},
  \citenamefont {Herman}, \citenamefont {Pisoni}, \citenamefont {Overweg},
  \citenamefont {Lee}, \citenamefont {Rickhaus}, \citenamefont {Watanabe},
  \citenamefont {Taniguchi}, \citenamefont {Sigrist}, \citenamefont {Ihn},\
  and\ \citenamefont {Ensslin}}]{Eich2018}%
  \BibitemOpen
  \bibfield  {author} {\bibinfo {author} {\bibfnamefont {M.}~\bibnamefont
  {Eich}}, \bibinfo {author} {\bibfnamefont {F.}~\bibnamefont {Herman}},
  \bibinfo {author} {\bibfnamefont {R.}~\bibnamefont {Pisoni}}, \bibinfo
  {author} {\bibfnamefont {H.}~\bibnamefont {Overweg}}, \bibinfo {author}
  {\bibfnamefont {Y.}~\bibnamefont {Lee}}, \bibinfo {author} {\bibfnamefont
  {P.}~\bibnamefont {Rickhaus}}, \bibinfo {author} {\bibfnamefont
  {K.}~\bibnamefont {Watanabe}}, \bibinfo {author} {\bibfnamefont
  {T.}~\bibnamefont {Taniguchi}}, \bibinfo {author} {\bibfnamefont
  {M.}~\bibnamefont {Sigrist}}, \bibinfo {author} {\bibfnamefont
  {T.}~\bibnamefont {Ihn}}, \ and\ \bibinfo {author} {\bibfnamefont
  {K.}~\bibnamefont {Ensslin}},\ }\href@noop {} {\bibfield  {journal} {\bibinfo
   {journal} {arXiv:1803.02923 [cond-mat]}\ } (\bibinfo {year} {2018})},\
  \Eprint {http://arxiv.org/abs/1803.02923} {arXiv:1803.02923 [cond-mat]}
  \BibitemShut {NoStop}%
\bibitem [{\citenamefont {Yan}\ and\ \citenamefont {Fuhrer}(2010)}]{Yan2010}%
  \BibitemOpen
  \bibfield  {author} {\bibinfo {author} {\bibfnamefont {J.}~\bibnamefont
  {Yan}}\ and\ \bibinfo {author} {\bibfnamefont {M.~S.}\ \bibnamefont
  {Fuhrer}},\ }\href {\doibase 10.1021/nl102459t} {\bibfield  {journal}
  {\bibinfo  {journal} {Nano Letters}\ }\textbf {\bibinfo {volume} {10}},\
  \bibinfo {pages} {4521} (\bibinfo {year} {2010})}\BibitemShut {NoStop}%
\bibitem [{\citenamefont {Avsar}\ \emph {et~al.}(2016)\citenamefont {Avsar},
  \citenamefont {{Vera-Marun}}, \citenamefont {Tan}, \citenamefont {Koon},
  \citenamefont {Watanabe}, \citenamefont {Taniguchi}, \citenamefont {Adam},\
  and\ \citenamefont {\"Ozyilmaz}}]{Avsar2016}%
  \BibitemOpen
  \bibfield  {author} {\bibinfo {author} {\bibfnamefont {A.}~\bibnamefont
  {Avsar}}, \bibinfo {author} {\bibfnamefont {I.~J.}\ \bibnamefont
  {{Vera-Marun}}}, \bibinfo {author} {\bibfnamefont {J.~Y.}\ \bibnamefont
  {Tan}}, \bibinfo {author} {\bibfnamefont {G.~K.~W.}\ \bibnamefont {Koon}},
  \bibinfo {author} {\bibfnamefont {K.}~\bibnamefont {Watanabe}}, \bibinfo
  {author} {\bibfnamefont {T.}~\bibnamefont {Taniguchi}}, \bibinfo {author}
  {\bibfnamefont {S.}~\bibnamefont {Adam}}, \ and\ \bibinfo {author}
  {\bibfnamefont {B.}~\bibnamefont {\"Ozyilmaz}},\ }\href@noop {} {\bibfield
  {journal} {\bibinfo  {journal} {Npg Asia Materials}\ }\textbf {\bibinfo
  {volume} {8}},\ \bibinfo {pages} {e274} (\bibinfo {year} {2016})}\BibitemShut
  {NoStop}%
\bibitem [{\citenamefont {Xiao}\ \emph {et~al.}(2010)\citenamefont {Xiao},
  \citenamefont {Chang},\ and\ \citenamefont {Niu}}]{Xiao2010}%
  \BibitemOpen
  \bibfield  {author} {\bibinfo {author} {\bibfnamefont {D.}~\bibnamefont
  {Xiao}}, \bibinfo {author} {\bibfnamefont {M.-C.}\ \bibnamefont {Chang}}, \
  and\ \bibinfo {author} {\bibfnamefont {Q.}~\bibnamefont {Niu}},\ }\href
  {\doibase 10.1103/RevModPhys.82.1959} {\bibfield  {journal} {\bibinfo
  {journal} {Reviews of Modern Physics}\ }\textbf {\bibinfo {volume} {82}},\
  \bibinfo {pages} {1959} (\bibinfo {year} {2010})}\BibitemShut {NoStop}%
\bibitem [{\citenamefont {Chang}\ and\ \citenamefont {Niu}(1996)}]{Chang1996a}%
  \BibitemOpen
  \bibfield  {author} {\bibinfo {author} {\bibfnamefont {M.-C.}\ \bibnamefont
  {Chang}}\ and\ \bibinfo {author} {\bibfnamefont {Q.}~\bibnamefont {Niu}},\
  }\href {\doibase 10.1103/PhysRevB.53.7010} {\bibfield  {journal} {\bibinfo
  {journal} {Physical Review B}\ }\textbf {\bibinfo {volume} {53}},\ \bibinfo
  {pages} {7010} (\bibinfo {year} {1996})}\BibitemShut {NoStop}%
\bibitem [{\citenamefont {McCann}\ \emph {et~al.}(2007)\citenamefont {McCann},
  \citenamefont {Abergel},\ and\ \citenamefont {Fal'ko}}]{McCann2007}%
  \BibitemOpen
  \bibfield  {author} {\bibinfo {author} {\bibfnamefont {E.}~\bibnamefont
  {McCann}}, \bibinfo {author} {\bibfnamefont {D.~S.}\ \bibnamefont {Abergel}},
  \ and\ \bibinfo {author} {\bibfnamefont {V.~I.}\ \bibnamefont {Fal'ko}},\
  }\href {\doibase 10.1140/epjst/e2007-00229-1} {\bibfield  {journal} {\bibinfo
   {journal} {The European Physical Journal Special Topics}\ }\textbf {\bibinfo
  {volume} {148}},\ \bibinfo {pages} {91} (\bibinfo {year} {2007})}\BibitemShut
  {NoStop}%
\bibitem [{\citenamefont {McCann}\ and\ \citenamefont
  {Koshino}(2013)}]{McCann2013}%
  \BibitemOpen
  \bibfield  {author} {\bibinfo {author} {\bibfnamefont {E.}~\bibnamefont
  {McCann}}\ and\ \bibinfo {author} {\bibfnamefont {M.}~\bibnamefont
  {Koshino}},\ }\href {\doibase 10.1088/0034-4885/76/5/056503} {\bibfield
  {journal} {\bibinfo  {journal} {Reports on Progress in Physics}\ }\textbf
  {\bibinfo {volume} {76}},\ \bibinfo {pages} {056503} (\bibinfo {year}
  {2013})}\BibitemShut {NoStop}%
\bibitem [{\citenamefont {Park}(2017)}]{Park2017}%
  \BibitemOpen
  \bibfield  {author} {\bibinfo {author} {\bibfnamefont {C.-S.}\ \bibnamefont
  {Park}},\ }\href {\doibase 10.1016/j.physleta.2017.10.044} {\bibfield
  {journal} {\bibinfo  {journal} {Physics Letters A}\ }\textbf {\bibinfo
  {volume} {382}} (\bibinfo {year} {2017}),\
  10.1016/j.physleta.2017.10.044}\BibitemShut {NoStop}%
\bibitem [{\citenamefont {Fuchs}\ \emph {et~al.}(2010)\citenamefont {Fuchs},
  \citenamefont {Pi\'echon}, \citenamefont {Goerbig},\ and\ \citenamefont
  {Montambaux}}]{Fuchs2010}%
  \BibitemOpen
  \bibfield  {author} {\bibinfo {author} {\bibfnamefont {J.~N.}\ \bibnamefont
  {Fuchs}}, \bibinfo {author} {\bibfnamefont {F.}~\bibnamefont {Pi\'echon}},
  \bibinfo {author} {\bibfnamefont {M.~O.}\ \bibnamefont {Goerbig}}, \ and\
  \bibinfo {author} {\bibfnamefont {G.}~\bibnamefont {Montambaux}},\ }\href
  {\doibase 10.1140/epjb/e2010-00259-2} {\bibfield  {journal} {\bibinfo
  {journal} {The European Physical Journal B}\ }\textbf {\bibinfo {volume}
  {77}},\ \bibinfo {pages} {351} (\bibinfo {year} {2010})}\BibitemShut
  {NoStop}%
\bibitem [{\citenamefont {Martin}\ \emph {et~al.}(2008)\citenamefont {Martin},
  \citenamefont {Blanter},\ and\ \citenamefont {Morpurgo}}]{Martin2008}%
  \BibitemOpen
  \bibfield  {author} {\bibinfo {author} {\bibfnamefont {I.}~\bibnamefont
  {Martin}}, \bibinfo {author} {\bibfnamefont {Y.~M.}\ \bibnamefont {Blanter}},
  \ and\ \bibinfo {author} {\bibfnamefont {A.~F.}\ \bibnamefont {Morpurgo}},\
  }\href {\doibase 10.1103/PhysRevLett.100.036804} {\bibfield  {journal}
  {\bibinfo  {journal} {Physical Review Letters}\ }\textbf {\bibinfo {volume}
  {100}},\ \bibinfo {pages} {036804} (\bibinfo {year} {2008})}\BibitemShut
  {NoStop}%
\bibitem [{\citenamefont {Cosma}\ and\ \citenamefont
  {Fal'ko}(2015)}]{Cosma2015}%
  \BibitemOpen
  \bibfield  {author} {\bibinfo {author} {\bibfnamefont {D.~A.}\ \bibnamefont
  {Cosma}}\ and\ \bibinfo {author} {\bibfnamefont {V.~I.}\ \bibnamefont
  {Fal'ko}},\ }\href {\doibase 10.1103/PhysRevB.92.165412} {\bibfield
  {journal} {\bibinfo  {journal} {Physical Review B}\ }\textbf {\bibinfo
  {volume} {92}},\ \bibinfo {pages} {165412} (\bibinfo {year}
  {2015})}\BibitemShut {NoStop}%
\bibitem [{\citenamefont {Xiao}\ \emph {et~al.}(2007)\citenamefont {Xiao},
  \citenamefont {Yao},\ and\ \citenamefont {Niu}}]{Xiao2007}%
  \BibitemOpen
  \bibfield  {author} {\bibinfo {author} {\bibfnamefont {D.}~\bibnamefont
  {Xiao}}, \bibinfo {author} {\bibfnamefont {W.}~\bibnamefont {Yao}}, \ and\
  \bibinfo {author} {\bibfnamefont {Q.}~\bibnamefont {Niu}},\ }\href {\doibase
  10.1103/PhysRevLett.99.236809} {\bibfield  {journal} {\bibinfo  {journal}
  {Physical Review Letters}\ }\textbf {\bibinfo {volume} {99}},\ \bibinfo
  {pages} {236809} (\bibinfo {year} {2007})}\BibitemShut {NoStop}%
\bibitem [{\citenamefont {Li}(2014)}]{Li2014}%
  \BibitemOpen
  \bibfield  {author} {\bibinfo {author} {\bibfnamefont {X.}~\bibnamefont
  {Li}},\ }\emph {\bibinfo {title} {Quantum {{Hall}} Effects in Novel {{2D}}
  Electron Systems : Nontrivial {{Fermi}} Surface Topology and Quantum {{Hall}}
  Ferromagnetism}},\ \href@noop {} {\bibinfo {type} {Dissertation thesis, the
  university of texas at austin}} (\bibinfo {year} {2014})\BibitemShut
  {NoStop}%
\bibitem [{\citenamefont {Cheianov}\ \emph {et~al.}(2012)\citenamefont
  {Cheianov}, \citenamefont {Aleiner},\ and\ \citenamefont
  {Fal'ko}}]{Cheianov1012}%
  \BibitemOpen
  \bibfield  {author} {\bibinfo {author} {\bibfnamefont {V.~V.}\ \bibnamefont
  {Cheianov}}, \bibinfo {author} {\bibfnamefont {I.~L.}\ \bibnamefont
  {Aleiner}}, \ and\ \bibinfo {author} {\bibfnamefont {V.~I.}\ \bibnamefont
  {Fal'ko}},\ }\href {\doibase 10.1103/PhysRevLett.109.106801} {\bibfield
  {journal} {\bibinfo  {journal} {Phys. Rev. Lett.}\ }\textbf {\bibinfo
  {volume} {109}},\ \bibinfo {pages} {106801} (\bibinfo {year}
  {2012})}\BibitemShut {NoStop}%
\bibitem [{\citenamefont {Thomas}\ \emph {et~al.}(1996)\citenamefont {Thomas},
  \citenamefont {Nicholls}, \citenamefont {Simmons}, \citenamefont {Pepper},
  \citenamefont {Mace},\ and\ \citenamefont {Ritchie}}]{Thomas1996}%
  \BibitemOpen
  \bibfield  {author} {\bibinfo {author} {\bibfnamefont {K.~J.}\ \bibnamefont
  {Thomas}}, \bibinfo {author} {\bibfnamefont {J.~T.}\ \bibnamefont
  {Nicholls}}, \bibinfo {author} {\bibfnamefont {M.~Y.}\ \bibnamefont
  {Simmons}}, \bibinfo {author} {\bibfnamefont {M.}~\bibnamefont {Pepper}},
  \bibinfo {author} {\bibfnamefont {D.~R.}\ \bibnamefont {Mace}}, \ and\
  \bibinfo {author} {\bibfnamefont {D.~A.}\ \bibnamefont {Ritchie}},\ }\href
  {\doibase 10.1103/PhysRevLett.77.135} {\bibfield  {journal} {\bibinfo
  {journal} {Phys. Rev. Lett.}\ }\textbf {\bibinfo {volume} {77}},\ \bibinfo
  {pages} {135} (\bibinfo {year} {1996})}\BibitemShut {NoStop}%
\bibitem [{\citenamefont {Thomas}\ \emph {et~al.}(1998)\citenamefont {Thomas},
  \citenamefont {Nicholls}, \citenamefont {Appleyard}, \citenamefont {Simmons},
  \citenamefont {Pepper}, \citenamefont {Mace}, \citenamefont {Tribe},\ and\
  \citenamefont {Ritchie}}]{Thomas1998}%
  \BibitemOpen
  \bibfield  {author} {\bibinfo {author} {\bibfnamefont {K.~J.}\ \bibnamefont
  {Thomas}}, \bibinfo {author} {\bibfnamefont {J.~T.}\ \bibnamefont
  {Nicholls}}, \bibinfo {author} {\bibfnamefont {N.~J.}\ \bibnamefont
  {Appleyard}}, \bibinfo {author} {\bibfnamefont {M.~Y.}\ \bibnamefont
  {Simmons}}, \bibinfo {author} {\bibfnamefont {M.}~\bibnamefont {Pepper}},
  \bibinfo {author} {\bibfnamefont {D.~R.}\ \bibnamefont {Mace}}, \bibinfo
  {author} {\bibfnamefont {W.~R.}\ \bibnamefont {Tribe}}, \ and\ \bibinfo
  {author} {\bibfnamefont {D.~A.}\ \bibnamefont {Ritchie}},\ }\href {\doibase
  10.1103/PhysRevB.58.4846} {\bibfield  {journal} {\bibinfo  {journal} {Phys.
  Rev. B}\ }\textbf {\bibinfo {volume} {58}},\ \bibinfo {pages} {4846}
  (\bibinfo {year} {1998})}\BibitemShut {NoStop}%
\bibitem [{\citenamefont {Micolich}(2011)}]{Micolich2011}%
  \BibitemOpen
  \bibfield  {author} {\bibinfo {author} {\bibfnamefont {A.~P.}\ \bibnamefont
  {Micolich}},\ }\href {http://stacks.iop.org/0953-8984/23/i=44/a=443201}
  {\bibfield  {journal} {\bibinfo  {journal} {Journal of Physics: Condensed
  Matter}\ }\textbf {\bibinfo {volume} {23}},\ \bibinfo {pages} {443201}
  (\bibinfo {year} {2011})}\BibitemShut {NoStop}%
\bibitem [{\citenamefont {Hamilton}\ \emph {et~al.}(2008)\citenamefont
  {Hamilton}, \citenamefont {Danneau}, \citenamefont {Klochan}, \citenamefont
  {Clarke}, \citenamefont {Micolich}, \citenamefont {Ho}, \citenamefont
  {Simmons}, \citenamefont {Ritchie}, \citenamefont {Pepper}, \citenamefont
  {Muraki},\ and\ \citenamefont {Hirayama}}]{Hamilton2008}%
  \BibitemOpen
  \bibfield  {author} {\bibinfo {author} {\bibfnamefont {A.~R.}\ \bibnamefont
  {Hamilton}}, \bibinfo {author} {\bibfnamefont {R.}~\bibnamefont {Danneau}},
  \bibinfo {author} {\bibfnamefont {O.}~\bibnamefont {Klochan}}, \bibinfo
  {author} {\bibfnamefont {W.~R.}\ \bibnamefont {Clarke}}, \bibinfo {author}
  {\bibfnamefont {A.~P.}\ \bibnamefont {Micolich}}, \bibinfo {author}
  {\bibfnamefont {L.~H.}\ \bibnamefont {Ho}}, \bibinfo {author} {\bibfnamefont
  {M.~Y.}\ \bibnamefont {Simmons}}, \bibinfo {author} {\bibfnamefont {D.~A.}\
  \bibnamefont {Ritchie}}, \bibinfo {author} {\bibfnamefont {M.}~\bibnamefont
  {Pepper}}, \bibinfo {author} {\bibfnamefont {K.}~\bibnamefont {Muraki}}, \
  and\ \bibinfo {author} {\bibfnamefont {Y.}~\bibnamefont {Hirayama}},\ }\href
  {http://stacks.iop.org/0953-8984/20/i=16/a=164205} {\bibfield  {journal}
  {\bibinfo  {journal} {Journal of Physics: Condensed Matter}\ }\textbf
  {\bibinfo {volume} {20}},\ \bibinfo {pages} {164205} (\bibinfo {year}
  {2008})}\BibitemShut {NoStop}%
\end{thebibliography}%

\end{document}